\documentclass[fleqn,usenatbib]{mnras}

\usepackage{newtxtext,newtxmath}

\usepackage[T1]{fontenc}

\DeclareRobustCommand{\VAN}[3]{#2}
\let\VANthebibliography\thebibliography
\def\thebibliography{\DeclareRobustCommand{\VAN}[3]{##3}\VANthebibliography}

\usepackage{graphicx}	
\usepackage{amsmath}	
\usepackage{siunitx}    
\usepackage{array}
\usepackage{multirow}
\usepackage{booktabs}

\newcommand{\borus}{\textsc{borus}}
\newcommand{\gro}{GRO J1655-40}
\newcommand{\gx}{GX 339-4}

\newcommand{\mbh}{$M_\mathrm{BH}$}
\newcommand{\mcol}[1]{\multicolumn{1}{c}{#1}}
\newcommand{\msol}{$\mathrm{M}_\odot$}

\newcommand{\nbmc}{$N_\mathrm{BMC}$}
\newcommand{\ngc}{NGC~4151}

\newcommand{\nustar}{{\it NuSTAR}}
\newcommand{\veldis}{$\sigma_\star$}
\newcommand{\xmm}{{\it XMM-Newton}}
\newcommand{\xspec}{\textsc{xspec}}
\newcommand{\xte}{XTE J1550-564}

\title[Assessing indirect methods]{Assessing indirect methods to determine black hole masses using NGC 4151}

\author[Williams et al.]{
James K. Williams,$^{1}$\thanks{E-mail: jwilli91@gmu.edu}
Mario Gliozzi,$^{1}$
Kyle A. Bockwoldt,$^{1}$
and Onic I. Shuvo$^{1}$
\\
$^{1}$Department of Physics and Astronomy, George Mason University, 4400 University Drive, Fairfax, VA 22030, USA
}

\date{Accepted XXX. Received YYY; in original form ZZZ}

\pubyear{2022}

\begin{document}
\label{firstpage}
\pagerange{\pageref{firstpage}--\pageref{lastpage}}
\maketitle

\begin{abstract}
Accurately determining the black hole mass (\mbh) in active galactic nuclei (AGN) is crucial to constraining their  properties and to studying their evolution. While direct methods yield reliable measurements of \mbh\ in unobscured type 1 AGN, where the dynamics of stellar or gas components can be directly observed, only indirect methods can be applied to the vast majority of heavily absorbed type 2 AGN, which represent most of the AGN population. Since it is difficult to evaluate the accuracy and precision of these indirect methods, we utilize the nearby X-ray bright Seyfert galaxy \ngc, whose \mbh\ has been tightly constrained with several independent direct methods, as a laboratory to assess the reliability of three indirect methods that have been applied to obscured AGN. All three, the X-ray scaling method, the fundamental plane of black hole activity, and the M--$\sigma$ correlation, yield \mbh\ values consistent with those inferred from direct methods and can therefore be considered accurate. However, only the X-ray scaling method and the M--$\sigma$ correlation are precise because the substantial scatter in the fundamental plane of BH activity allows only for crude estimates. Of the four M--$\sigma$ correlations we used, only the one from Kormendy and Ho yields a value consistent with the dynamical estimates. This study suggests that the best approach to estimating the black hole mass in systems where direct dynamical methods cannot be applied is to utilize a combination of indirect methods, taking into account their different ranges of applicability.
\end{abstract}




\section{Introduction}
Only a few decades ago the very existence of black holes (BHs) was still debated. Now it is widely accepted that stellar-mass BHs (with typical \mbh\ values in the 3--20 M$_\odot$ range) are routinely observed in X-ray binaries in our Galaxy, and supermassive black holes (SMBHs; $10^6\textrm{--}10^9$ M$_\odot$) probably exist at the centers of all massive galaxies. Of the three physical properties that completely characterize a BH (mass, spin, and electric charge), the black hole mass \mbh\ plays a major role by defining the length and time scales around BHs as well as constraining the accretion rate via the Eddington ratio $\lambda_\mathrm{Edd}=L_\mathrm{bol}/L_\mathrm{Edd}$ (i.e., the ratio between the bolometric luminosity and the Eddington luminosity $L_\mathrm{Edd}=1.3\times 10^{38} (M_\mathrm{BH}/\mathrm{M_\odot})$ erg s$^{-1}$).

Importantly, in an SMBH the mass reveals crucial information about the BH's growth over cosmic times and makes possible a comparison with the host galaxy evolution, suggesting, at least for galaxies with bulges, a possible coevolution of SMBHs and their galaxies (see \citealt{Kormendy2013} for a review). It is therefore of paramount importance to constrain the mass of SMBHs at all cosmological stages, in different spatial environments, and in different types of active galactic nuclei (AGN).

The most reliable way to determine \mbh\ is to directly measure the dynamics of a `test particle' in the BH's sphere of influence, which is defined by the radius $r_\mathrm{infl}=G M_\mathrm{BH}/\sigma^2$, where $\sigma$ is the stellar velocity dispersion of the host bulge. In stellar binary systems this is accomplished by measuring the orbital parameters of a visible companion star. Similarly, in our Galaxy \mbh\ was accurately measured by monitoring the orbits of the innermost stars over more than a decade \citep{Ghez2008, Genzel2010}. For quiescent or weakly active galaxies, \mbh\ can be obtained by the observation and modeling of stellar or gas components resolved in the BH's sphere of influence (e.g., \citealt{Gebhardt2011, Macchetto1997}). In bright type 1 AGN, \mbh\ can be directly determined by measuring the dynamics of the broad line region (BLR) via the reverberation mapping technique (e.g., \citealt{Peterson2004}). In type 2 AGN, where the BLR is not accessible to observations, the dynamical estimate of \mbh\ is possible only for a handful of objects, thanks to the measurement of megamasers in their Keplerian disk (e.g., \citealt{Kuo2011}). However, for the vast majority of type 2 AGN, it is necessary to rely on indirect methods to constrain \mbh. 

Since type 2 AGN represent the majority of AGN and since the SMBH growth is expected to occur when the AGN are deeply buried in gas and dust at the center of galaxies possibly merging (e.g., \citealt{Blecha2018}), having reliable methods to accurately constrain \mbh\ in obscured AGN is crucial to understanding their evolution over cosmic time. In this work, we will use \ngc, a nearby, bright type 1 AGN, whose \mbh\ has been determined with many independent direct methods, as a laboratory to test the accuracy and reliability of three indirect methods that can be used for type 2 AGN: the X-ray scaling method \citep{Shaposhnikov2009, Gliozzi2011}, the fundamental plane of BH activity \citep{Merloni2003, Gueltekin2019}, and the M--$\sigma$ correlation \citep{McConnell2013, Woo2013}. 

\begin{figure}
 \includegraphics[width=\columnwidth]{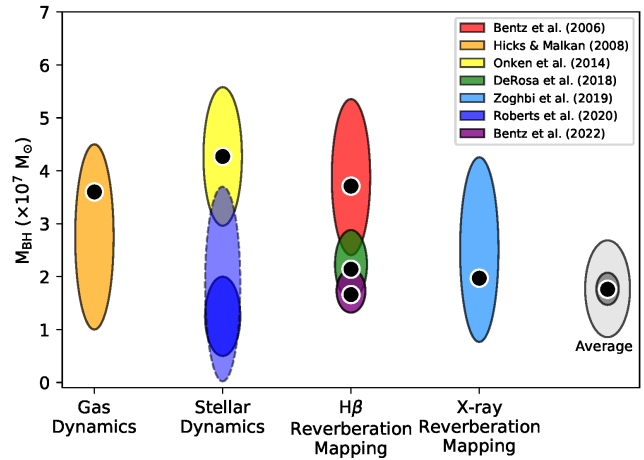}
 \caption{Visual comparison of the \mbh\ estimates obtained with different direct methods. The \mbh\ values are indicated by the darker circles within the ellipses whose vertical size represents the uncertainty. For consistency, the values obtained from gas and stellar dynamics were rescaled using the same distance of 15.8 Mpc based on Cepheid measurements \citep{Yuan2020}, and reverberation mapping values were rescaled using the same average virial scaling factor of 4.47 \citep{Woo2015}. The average (in gray) was obtained by taking the weighted mean of all the independent direct methods, and the ellipses around it represent the 1$\sigma$ and 3$\sigma$ confidence levels.}
 \label{fig:directMBH}
\end{figure}

Before focusing on the comparison of these indirect methods, we summarize the results obtained from the direct methods in Table~\ref{tab:direct_methods} and in Fig.~\ref{fig:directMBH}, revealing a general agreement within the respective uncertainties. For consistency, for all methods based on the source distance, we used the same value, $D=15.8\pm0.4$ Mpc, obtained from Cepheid measurements \citep{Yuan2020}, and for the virial factor, which accounts for the unknown geometry and kinematics of the BLR in reverberation mapping methods, we used the average value ${\langle f\rangle}=4.47\pm1.25$ obtained by \citet{Woo2015}. Rescaling with this distance, the \mbh\ obtained by \citet{Hicks2008} via gas dynamics is $3.6\times 10^7~\mathrm{M_\odot}$, and the one obtained by \citet{Onken2014} using stellar dynamics is $4.27\times 10^7~\mathrm{M_\odot}$. With the same technique but with higher-quality data and refined modeling methods, \citet{Roberts2021} constrained the \mbh\ range to be $5\times10^6-2\times 10^7~\mathrm{M_\odot}$, which they extended to $2.5\times10^6-3\times 10^7~\mathrm{M_\odot}$ to be more conservative. Using H$_\beta$ reverberation mapping, \citet{Bentz2006} derived $3.71\times 10^7~\mathrm{M_\odot}$. With the same technique but higher-quality data, \citet{Derosa2018} obtained instead $2.14\times 10^7~\mathrm{M_\odot}$. Very recently, by directly modeling the BLR from the reverberation mapping campaign (and, hence, without making any assumption on the virial factor $f)$, \citet{Bentz2022} obtained  $1.66\times10^7~\mathrm{M_\odot}$. Finally, using X-ray reverberation mapping, where the time delay of the variable component of the narrow Fe K$\alpha$ line is measured with respect to the X-ray continuum, \citet{Zoghbi2019} inferred $1.92\times 10^7~\mathrm{M_\odot}$.

\begin{table*}
	\centering
	\caption{\mbh\ measurements derived from direct methods}
	\label{tab:direct_methods}
	\begin{tabular}{llccc}
		\hline
        \hline
		Reference & Method & \mbh\ & \mbh\ min & \mbh\ max \\
		& & (\msol) & (\msol) & (\msol) \\
		\hline
		\noalign{\smallskip}
		\citet{Hicks2008} & gas dynamics & $3.6 \times 10^7$ & $1.0 \times 10^7$ & $4.5 \times 10^7$ \\
		\citet{Onken2014} & stellar dynamics & $4.3 \times 10^7$ & $3.0 \times 10^7$ & $5.6 \times 10^7$ \\
		\cite{Roberts2021} & stellar dynamics & $\ldots$ & $2.5 \times 10^6$ & $3.0 \times 10^7$ \\
		\cite{Bentz2006} & H~$\beta$\ reverberation mapping & $3.7 \times 10^7$ & $2.4 \times 10^7$ & $5.4 \times 10^7$ \\
		\cite{Derosa2018} & H~$\beta$\ reverberation mapping & $2.1 \times 10^7$ & $1.6 \times 10^7$ & $2.9 \times 10^7$ \\
		\cite{Bentz2022} & H~$\beta$\ reverberation mapping & $1.7 \times 10^7$ & $1.3 \times 10^7$ & $2.1 \times 10^7$ \\
		\cite{Zoghbi2019} & X-ray reverberation mapping & $1.9 \times 10^7$ & $7.6 \times 10^6$ & $4.3 \times 10^7$ \\
		\hline
	\end{tabular}
	\begin{flushleft}
		Note that our values here do not exactly match those in the references because we modified them slightly for consistency. For all methods based on the source distance (gas and stellar dynamics), we used the same value $D=15.8\pm0.4$ Mpc obtained from Cepheid measurements \citep{Yuan2020}, and for the reverberation mapping values we used the average virial factor ${\langle f\rangle}=4.47\pm1.25$ obtained by \citet{Woo2015}.
	\end{flushleft}
\end{table*}

For the sake of simplicity and to directly quantify the comparison with the different indirect method estimates, we utilize the weighted mean obtained from all four different direct methods, where we discarded the oldest estimates from stellar dynamics \citep{Onken2014} and H$_\beta$ reverberation mapping \citep{Bentz2006}, which are superseded by more recent and accurate estimates obtained with the same methods. To compute the weighted mean, we first took the average of the minimum and maximum uncertainties in case they were asymmetric and then used the formula $(\Sigma x_i/\sigma_i^2)/(\Sigma(1/\sigma_i^2))$. For the 1$\sigma$ uncertainty on the weighted mean we utilized the formula $(1/(\Sigma(1/\sigma_i^2)))^{1/2}$.
This choice yields ${\langle M_{\mathrm BH}\rangle}=1.76\times 10^7~\mathrm{M_\odot}~ (\sigma=3.05\times10^6~\mathrm{M_\odot}$), which is illustrated in Fig.~\ref{fig:directMBH} with the 1$\sigma$ and 3$\sigma$ ellipses.

The structure of the paper is the following. In Section 2, we describe the X-ray data reduction. In Section 3, we report on the temporal and spectral analysis of \nustar\ data. The estimates of \mbh\ using three different indirect methods are described in Section 4. We summarize the main findings and draw our conclusions in Section 5. In the Appendix, we give a quick guide to the X-ray scaling method.

\section{X-ray Data Reduction}
\label{sec:datareduction}
In order to apply the X-ray scaling method to \ngc\ or to any other non-jet-dominated AGN, it is crucial to properly characterize the primary emission believed to be produced via Comptonization in the corona. Among the available X-ray observatories, \nustar, with its high sensitivity in the 3--79 keV energy range and low background, is the best instrument to tightly constrain the contributions of absorption and reflection and hence correctly measure the primary emission.

\ngc\ has been observed 12 times with \nustar. At the time of writing, one of those 12 observations was not available in the archive, and two of them had exposures too short to analyse (209~s and 17~s). Of the remaining nine, in this work we use the four observations in which the source had high count rates ($\ge 8.68$ c/s) and was not extremely variable ($F_{\mathrm{var,soft}} \le 7.7$ per cent and $F_{\mathrm{var,hard}} \le 6.6$ per cent). The choice of high count rate translates into a higher signal-to-noise ratio (S/N) for the spectral analysis and at the same time ensures the applicability of the X-ray scaling method, which cannot reliably be used for BH systems accreting at a very low rate \citep{Jang2014}. The second criterion of our selection ensures that the spectral variability is negligible and thus it is appropriate to carry out a time-averaged spectral analysis. We use four observations rather than just one to demonstrate that the results obtained from the X-ray scaling method are not dependent on a specific observation.

All nine observations are listed in Table~\ref{tab:obsids}. For the sake of brevity, we give the short names `Obs1' through `Obs4' to the four we analysed. We processed them with the \nustar\ data analysis pipeline \textsc{nupipeline}. Source photons were taken from a circle of radius 70 arcsec centered on the source and background photons from a circle of 70 arcsec nearby in the same image. We then extracted spectra, light curves, and images using the \textsc{nuproducts} commmand for both the FPMA and FPMB sensors on the observatory. There was no need to correct for pile-up because \nustar's pile-up window is so short that it is only a problem for extremely bright sources with count rates greater than 10,000 s$^{-1}$, much higher than the $\sim$9--11 s$^{-1}$ in our observations. The last column in Table~\ref{tab:obsids} lists the count rate for FPMA only. The rate for FPMB was within 0.5 count s$^{-1}$ of that for FPMA in every case (but consistently slightly smaller).
\begin{table*}
	\centering
	\caption{\nustar\ observations of \ngc} 
	\label{tab:obsids}
		\begin{tabular}{lccccr}
			\hline
			\hline
			\multirow{2}{*}{ObsID} & \multirow{2}{*}{Date} & \multirow{2}{*}{Exposure time} & \multicolumn{2}{c}{$F_{\mathrm{var}}$}
			& \multirow{2}{*}{Count rate} \\
			\cmidrule(lr){4-5}
			& & (s) & soft & hard & \multicolumn{1}{c}{(s$^{-1}$)}\\
			\hline
			\noalign{\smallskip}
			60001111002 (`Obs1') & 2012-11-12 & 21861 & $0.062 \pm 0.005$ & $0.065 \pm 0.006$ & $9.69 \pm 0.04$\\
			60001111003 & 2012-11-12 & 57034 & $0.090\pm0.003$ & $0.075\pm0.004$ & $8.92 \pm 0.02$\\
			60001111005 & 2012-11-14 & 61528 & $0.093\pm0.003$ & $0.089\pm0.004$ & $10.29 \pm 0.03$\\
			60502017002 & 2019-07-24 & 31738 & $0.134\pm0.007$ & $0.114\pm0.008$ & $8.88 \pm 0.04$\\
			60502017004 (`Obs2') & 2019-11-12 & 43738 & $0.058 \pm 0.003$ & $0.051 \pm 0.004$ & $11.48 \pm 0.04$\\
			60502017006 (`Obs3') & 2019-12-24 & 32401 & $0.052 \pm 0.003$ & $0.040 \pm 0.005$ & $10.90 \pm 0.03$\\
			60502017008 & 2020-01-10 & 30586 & $0.029\pm0.010$ & $0.072\pm0.010$ & $8.36 \pm 0.04$\\
			60502017010 & 2020-01-19 & 29717 & $0.100\pm0.008$ & $0.084\pm0.001$ & $8.84 \pm 0.05$\\
			60502017012 (`Obs4') & 2020-01-23 & 28859 & $0.077\pm0.004$ & $0.066\pm0.003$ & $8.68 \pm 0.03$\\
			\hline
		\end{tabular}
	\begin{flushleft}
		{$F_{\mathrm{var}}$ is the fractional variability parameter, defined in equation (1), which measures the intrinsic variability amplitude relative to the mean count rate, corrected for the effect of random errors. Here, soft is 3.5--10 keV and hard is 10--70 keV. The observations marked off with the names `Obs1' through `Obs4' are the ones analysed in this paper.}
	\end{flushleft}
\end{table*}

\section{X-ray Analysis}
\label{sec:Xanalysis}
\subsection{X-ray temporal analysis}
\label{sec:Xvariab}  
Because of its X-ray brightness \ngc\ has been the target of all the major X-ray telescopes in the last few decades. A recent reanalysis of all the available \xmm\ observations performed by \citet{Zoghbi2019} demonstrates that \ngc\ varies on many different timescales and that the origin of this variability can be ascribed to intrinsic changes in the primary coronal emission as well as variations of the complex absorption components along the line of sight.

We first analysed the light curves of the four observations extracted in the whole energy band from 3 to 79 keV, using both the FPMA and FPMB instruments. For all observations, the background light curves are very stable and the count-rate values are lower than 1 per cent of the source values; as a consequence, there is basically no difference between the background-subtracted source light curves and the ones not background-subtracted. Additionally, the source light curves of FPMA and FPMB are almost indistinguishable. Therefore, for the energy-selected temporal analysis, we focused on FPMA only and did not subtract the background.
\begin{figure*}
\includegraphics[bb=85 15 430 455,clip=,angle=0,width=8.6cm]{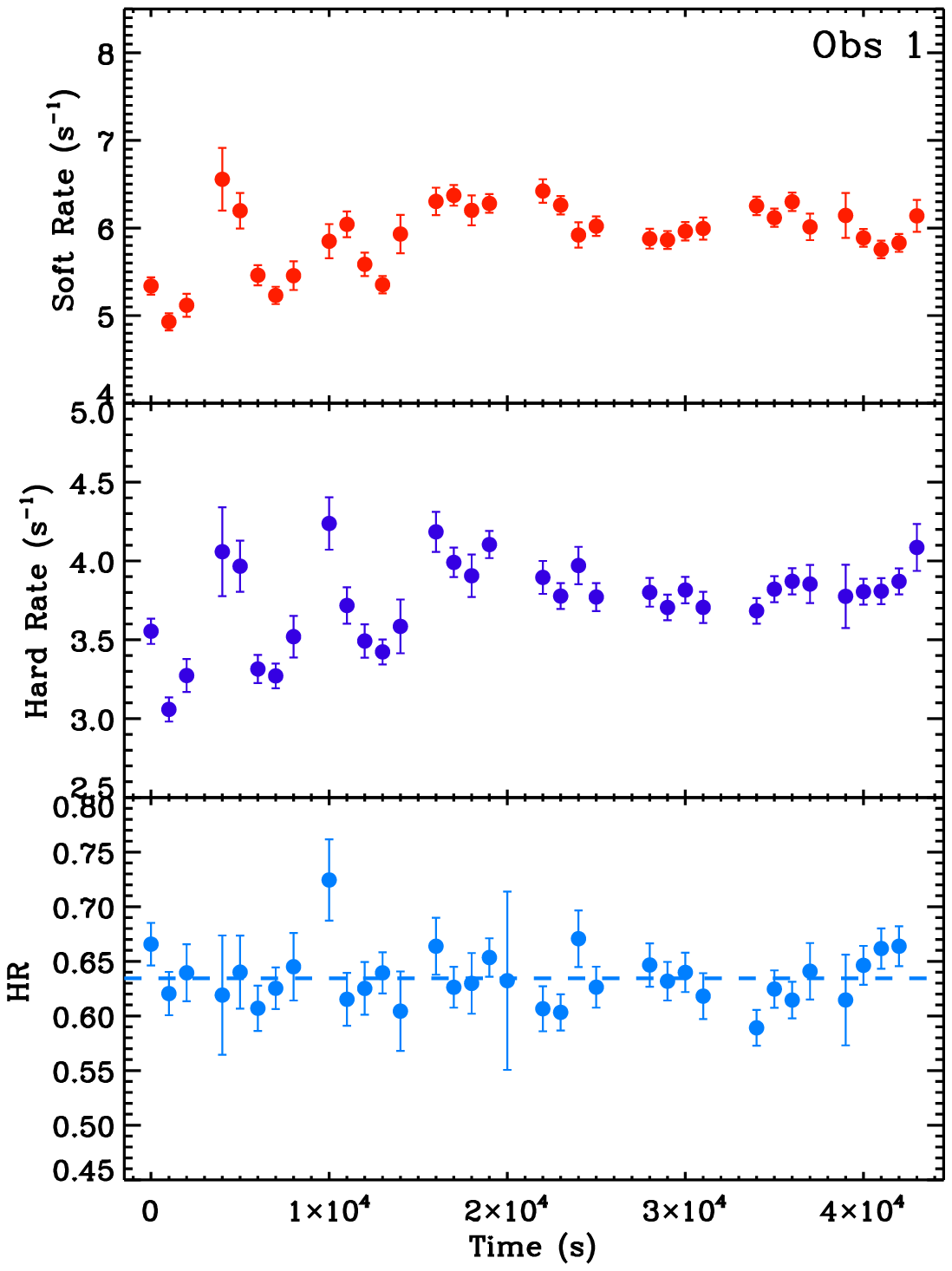}
\includegraphics[bb=85 15 430 455,clip=,angle=0,width=8.6cm]{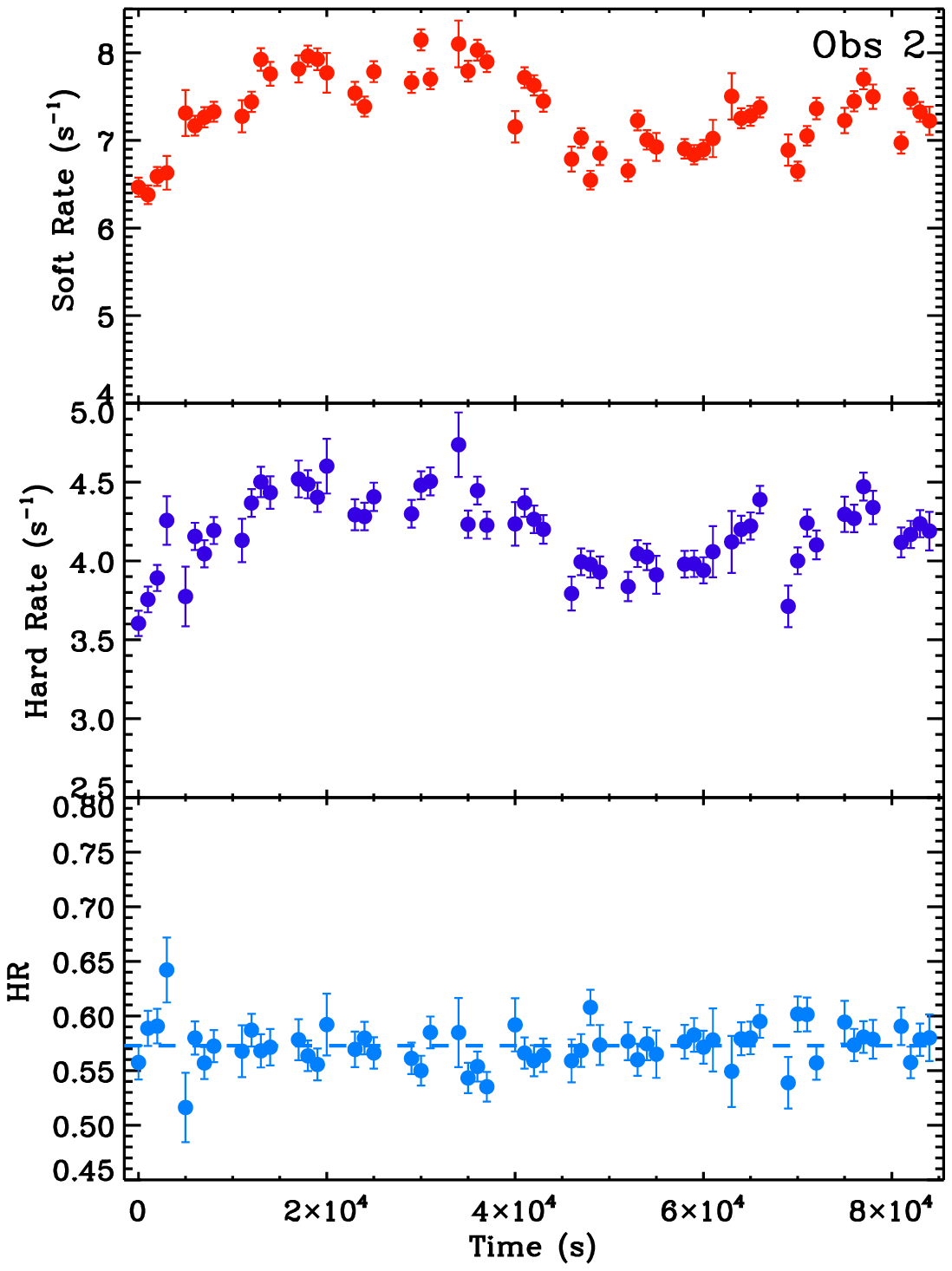}
\includegraphics[bb=85 15 430 455,clip=,angle=0,width=8.6cm]{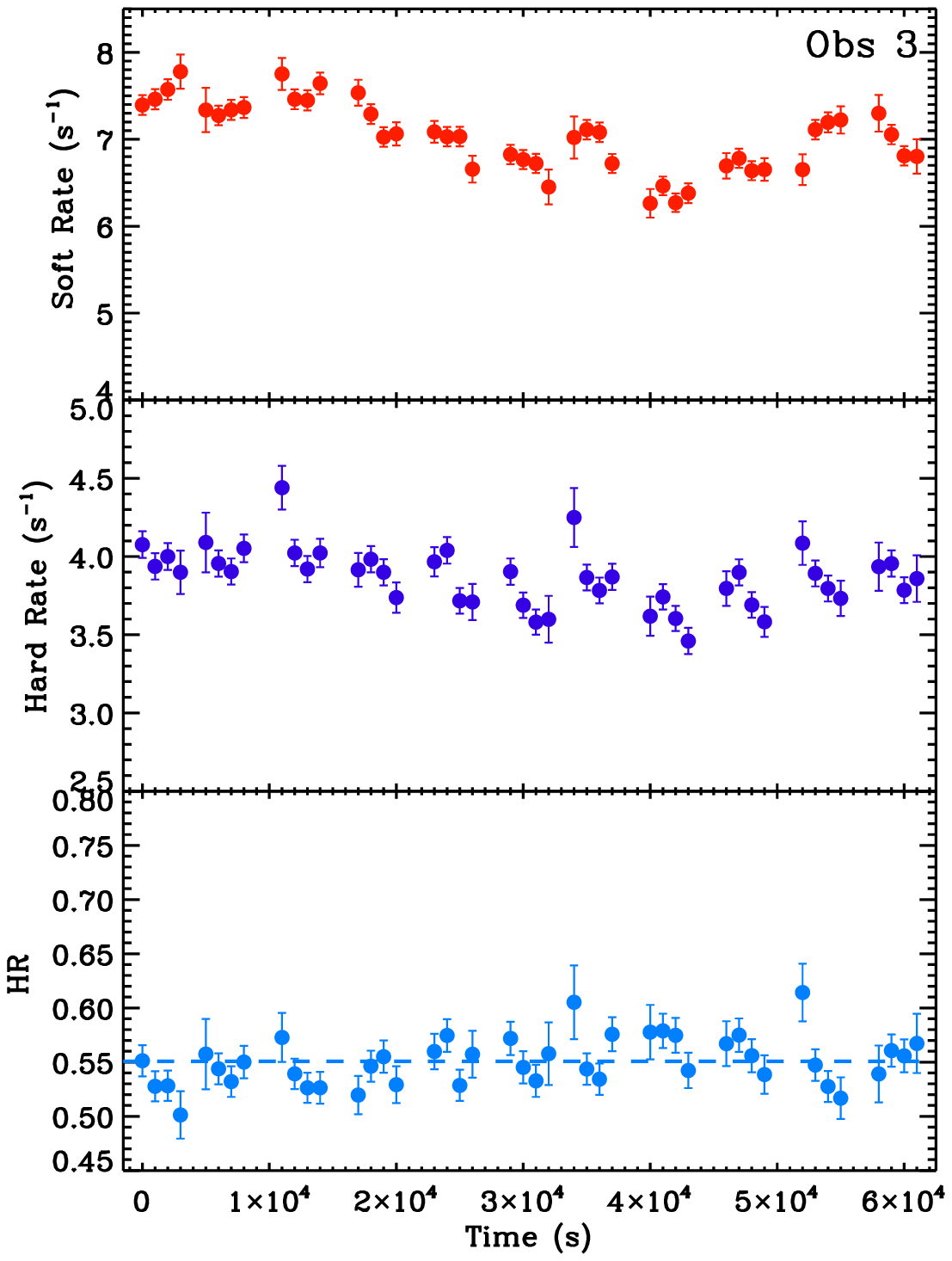}
\includegraphics[bb=85 15 430 455,clip=,angle=0,width=8.6cm]{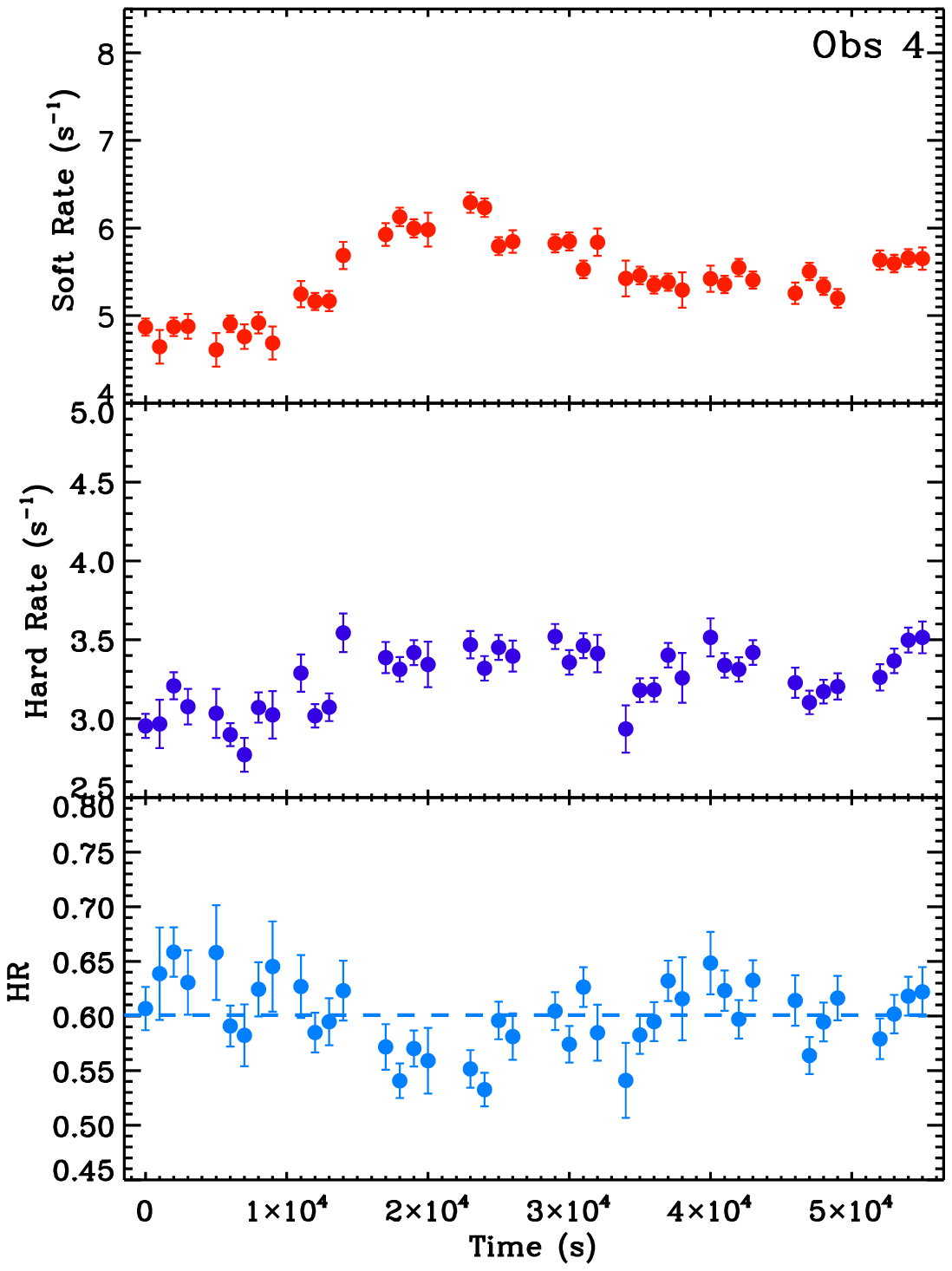}
\caption{Light curves of the soft (3.5--10 keV) count rates (top panels), hard (10--70 keV) count rates (middle panels), and hardness ratio (hard/soft, bottom panels) for the four sets of {\it NuSTAR} observations used in this work. Time bins are 1000 s.}
\label{fig:light_curves}
\end{figure*}

The soft (3.5--10 keV), hard (10--70 keV), and hardness ratio (HR = hard/soft) light curves of the four observations are illustrated in Fig.~\ref{fig:light_curves} and can be summarized as follow. For all observations, both soft and hard light curves show statistically significant variability (according to a $\chi^2$ test) with low-amplitude count-rate changes. To quantify the variability in these two bands and compare all observations, we used the fractional variability parameter $F_{\rm var}$, which measures the intrinsic variability amplitude relative to the mean count rate, corrected for the effect of random errors: 
\begin{equation} F_{\rm
var}=\frac{(\sigma^2-\Delta^2)^{1/2}}{\langle r\rangle} \end{equation} 
\noindent
where $\sigma^2$ is the variance, $\langle r\rangle$ the unweighted mean count
rate, and $\Delta^2$ the mean square value of the uncertainties associated with
each individual count rate. The values of $F_{\rm var,soft}$ and $F_{\rm var,hard}$ are reported in Table~\ref{tab:obsids}.

Since the count-rate variations in the soft and hard bands occur mostly in concert, the HR light curves are nearly flat and consistent with the hypothesis of constancy. Only Obs4 (bottom right panel of Fig.~\ref{fig:light_curves}) reveals some moderate spectral variability; however, only a few points of the HR light curve show a departure from the average value larger than 1$\sigma$. We can therefore conclude that the spectral variability in each of the four observations is negligible.

\subsection{X-ray spectral analysis}
The spectral analysis of \ngc\ has been the focus of several works, which have revealed a complex spectrum characterized by several components (e.g., \citealt{Yaqoob1989}). We base our spectral analysis on the results recently obtained by \citet{Zoghbi2019}, who reanalysed all the available \xmm\ data combined with three \nustar\ exposures and four {\it Suzaku} observations and obtained a best-fitting model comprising two ionized absorption models, a Bremsstrahlung with two Gaussian lines to characterize the soft part of the spectrum, and a partially absorbed power law plus a cold reflection component to model the hard part of the spectrum. 

Since we are interested only in constraining the coronal primary emission and since the \nustar\ energy range starts at 3 keV, we can disregard the soft X-ray components as well as the warm absorbers. Our baseline model in the \xspec\ syntax is the following:
\begin{verbatim}
phabs * (atable(borus) + zpcfabs*bmc)
\end{verbatim}
where the absorption model \textsc{phabs} accounts for our Galaxy contribution, the \borus\ table model parametrizes the continuum scattering and fluorescent emission line components associated with the torus, and \textsc{zpcfabs} models the neutral partial absorption acting on the primary emission, which is described by the Comptonization model \textsc{bmc}. To maintain the self-consistency of the physically motivated torus model \borus, the spectral index of the \textsc{bmc} model was linked to the photon index value of the torus model by the relationship $\alpha=\Gamma-1$ and the normalization of the Comptonization model \nbmc\ was forced to be equal to the \borus\ power-law normalization divided by a factor of 30, based on the empirical relationship obtained in \citet{Gliozzi2021}. Changing this factor of 30 by $\pm1\sigma$ has little impact on the spectral analysis and yields \mbh\ values that vary by less than 25 per cent (see table 3 in \citet{Gliozzi2021}).

\begin{figure*}
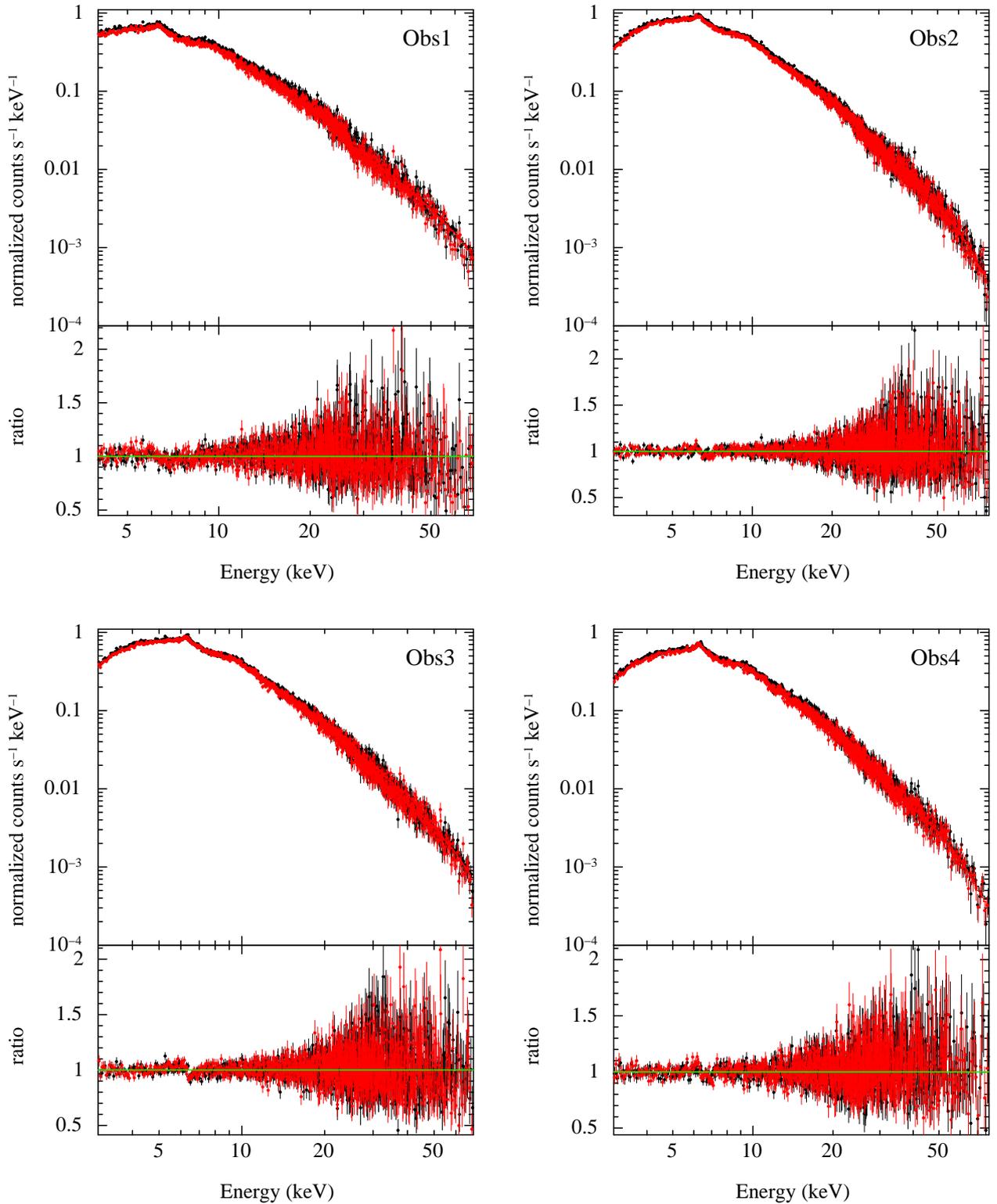

\includegraphics[width=8.6cm]{spec_ratio_Obs1.eps}
\includegraphics[width=8.6cm]{spec_ratio_Obs2.eps}
\includegraphics[width=8.6cm]{spec_ratio_Obs3.eps}
\includegraphics[width=8.6cm]{spec_ratio_Obs4.eps}
\caption{The top panels show the \nustar\ spectra of the four observations (black data points indicate FPMA data and red ones the FPMB data) with the best-fitting models overlaid; the bottom panels show the data-to-model ratios.}
\label{fig:spectra}
\end{figure*}

This model fits the four \nustar\ observations reasonably well, as shown by the results reported in Table~\ref{tab:spectral_results} and the best fits and model-to-data ratios shown in Fig.~\ref{fig:spectra}. For Obs1, the errors for both the covering factor CFtor and the covering fraction CvrFract were unconstrained, and therefore we do not include them in the table. Likewise, $\log A$ is generally poorly constrained but has relatively little effect on the \mbh\ calculation; therefore, we did not include its errors either. 

\begin{table*}
	\caption{Spectral results}
	\label{tab:spectral_results}
	\begin{center}
		\begin{tabular}{lcccccccccc} 
			\hline
			\hline    
			\noalign{\smallskip}
			\mcol{Observation} & \mcol{$\log(N_{{\textrm{H}}_{\textrm{tor}}})$} & \mcol{CFtor} & 
			\mcol{$N_{{\textrm{H}}_{\textrm{zpcfabs}}}$} & \mcol{CvrFract} & \mcol{$\Gamma$} & \mcol{\nbmc} & \mcol{kT} & \mcol{$\log{A}$} & \mcol{$\chi^2/$dof} & $L_\mathrm{2-10\ keV}$\\
			
			& & & \mcol{($10^{22}$ cm$^{-2}$)} & & & & \mcol{(keV)} & & & (erg s$^{-1}$)\\
			
			\mcol{(1)} & \mcol{(2)} & \mcol{(3)} & \mcol{(4)} & \mcol{(5)} & 
			\mcol{(6)} & \mcol{(7)} & \mcol{(8)} & \mcol{(9)} & \mcol{(10)} &
			\mcol{(11)}\\
			\hline
			\noalign{\smallskip}
			Obs1 & $24.04_{-0.02}^{+0.03}$ & $0.91$ & $18.6_{-0.3}^{+0.5}$ & $0.64$ & $1.83_{-0.01}^{+0.01}$ & $2.18_{-0.02}^{+0.02}\times10^{-3}$ & $0.11_{-0.04}^{+0.05}$ & $1.45$ & 1677.5/1563 & $8.26 \times 10^{42}$ \\
			\noalign{\smallskip}
			Obs2 & $24.10_{-0.01}^{+0.01}$ & $0.90_{-0.01}^{+0.02}$ & $13.9_{-0.3}^{+0.4}$ & $0.73_{-0.01}^{+0.01}$ & $1.87_{-0.01}^{+0.01}$ & $2.74_{-0.01}^{+0.01}\times10^{-3}$ & $0.12_{-0.01}^{+0.01}$ & $7.36$ & 2279.5/2076 & $1.07 \times 10^{43}$\\
			\noalign{\smallskip}
			Obs3 & $23.97_{-0.01}^{+0.01}$ & $0.90_{-0.02}^{+0.01}$ & $9.0_{-0.9}^{+0.5}$ & $0.74_{-0.03}^{+0.05}$ & $1.83_{-0.01}^{+0.01}$ & $2.40_{-0.01}^{+0.01}\times10^{-3}$ & $0.17_{-0.01}^{+0.01}$ & $1.45$ & 1999.5/1822 & $9.10 \times 10^{42}$ \\
			\noalign{\smallskip}
			Obs4 & $24.08_{-0.03}^{+0.03}$ & $0.80_{-0.02}^{+0.03}$ & $17.1_{-0.7}^{+0.7}$ & $0.69_{-0.01}^{+0.01}$ & $1.82_{-0.01}^{+0.01}$ & $2.08_{-0.1}^{+0.1}\times10^{-3}$ & $0.18_{-0.01}^{+0.01}$ & $1.45$ & 1749.2/1710 & $7.96 \times 10^{42}$ \\
			\noalign{\smallskip}
			\hline
		\end{tabular}
	\end{center}
	\begin{flushleft}
		Columns: 1 = observation. 2 = logarithm of the torus column density calculated by the \textsc{borus02} model. 3 = torus covering factor calculated by \textsc{borus02}. 4 = column density calculated by the \textsc{zpcfabs} model. 5 = covering fraction calculated by \textsc{zpcfabs}. 6 = photon index. 7 = normalization of the \textsc{bmc} model. 8 = temperature in keV. 9 = logarithm of $A$, where $f=A/(1+A)$ is the fraction of seed photons that are scattered. 10 = $\chi^2$ divided by degrees of freedom. 11 = X-ray luminosity in the band 2--10 keV.
	\end{flushleft}
\end{table*}

\label{sec:Xspectra}

\section{Black hole mass estimates with indirect methods}
\label{sec:MBH_estimate}
In the following subsections, we describe the basic characteristics and the application to \ngc\ of three indirect methods that can be used to constrain the black hole mass in type 2 AGN, where the innermost region (comprising the BLR and the sphere of influence of the BH) is not directly accessible and hence direct dynamical methods cannot be applied.
\subsection{X-ray scaling method}
\label{sec:Xscaling}
The X-ray scaling method was first introduced by \citet{Shaposhnikov2009} to constrain the black hole mass and distance in X-ray binary systems (XRBs). It was later successfully applied  to type 1 AGN, showing that the \mbh\ values obtained with this method were consistent with those determined via reverberation mapping \citep{Gliozzi2011}, and more recently extended to type 2 AGN, utilizing a sample of heavily obscured AGN with \mbh\ constrained via megamaser measurements \citep{Gliozzi2021}. In short, this method, which is based on the assumption that the spectral and temporal properties of the X-rays produced via Comptonization by the hot corona surrounding BHs are the same at all scales (see, e.g., \citealt{Done2005,Koerding2006,McHardy2006,Sobolewska2011,Ruan2019}), is able to determine the \mbh\ of any supermassive BH accreting at high or moderate level by scaling up the dynamically constrained \mbh\ of a stellar-mass BH, used as a reference. To this end, we utilize the following scaling equation:
\begin{equation}\label{eq_Xscal1}
M_{\mathrm{BH,AGN}}=M_{\mathrm{BH,ref}}  \times \left(\frac{N_{\mathrm{BMC,AGN}}}{N_{\mathrm{BMC,ref}}}\right)  \times \left(\frac{d_{\mathrm{AGN}}^2}{d_{\mathrm{ref}}^2}\right)
\end{equation}
where \nbmc\ is the normalization of the \textsc{bmc} model (measured in units of $10^{39}$ erg s$^{-1}$ divided by the distance squared in units of 10 kpc) and $d$ the distance in kpc. A derivation of this equation is provided in the Appendix. Note that this method cannot be applied to jet-dominated sources, since it is based on the assumption that the bulk of the X-rays are produced by the corona.

More specifically, this method exploits the positive correlation between the X-ray photon index $\Gamma$ and the accretion rate (which implies that $\Gamma$ is a reliable indicator of the accretion state in BHs accreting above 1 per cent of the Eddington limit), the so-called `softer when brighter' trend, which is regularly observed in XRB transitions between canonical spectral states as well as in AGN samples (e.g., \citealt{Shemmer2008,Brightman2013}) and individual AGN in long-term monitoring campaigns \citep{Sobolewska2009}.

Further details on this method and the basic steps necessary to quickly derive \mbh\ from the X-ray spectral results are provided in the Appendix. Here, for completeness, we report the \mbh\ values obtained using all the different reference stellar-mass BHs during spectral transitions introduced in \citet{Gliozzi2011}. As shown in our previous works on the reverberation mapping AGN sample and on the megamaser type 2 AGN sample, the most accurate \mbh\ values are obtained using the reference patterns of \gro\ during the spectral decay of the 2005 outburst (hereafter GROD05) and \gx\ during the spectral decay of the 2003 outburst (hereafter GXD03). Using the spectral trends during the outburst rise (XTER98, GROR05, GXR04) yields \mbh\ values that are systematically lower than the corresponding values obtained with dynamical methods. Nevertheless, the use of  XTER98 is crucial when dealing with steep photon indices, which are not covered by the GROD05 and GXD03 patterns (see the Appendix for more details). In that case, a reasonably good agreement with the \mbh\ values derived using GROD05 and GXD03 is obtained using the rising pattern of \xte\ during the 1998 outburst multiplied by a factor of 4 (hereafter 4*XTER98). We note that this is slightly different from the correcting factor of 3 inferred by \citet{Gliozzi2011} using the AGN reverberation mapping sample.

The results of this method using the different reference patterns and the four different \nustar\ observations are reported in Table~\ref{tab:Xscaling}. Since the spectral fits from the four \nustar\ observations are fairly similar, the \mbh\ values obtained from the different observations are nearly identical and consistent within their statistical uncertainties. For each reference pattern, we also report the average \mbh\ from the four observations, along with the the minimum and maximum values of \mbh, which take into account the statistical errors on the spectral parameters $\Gamma$ and \nbmc\ and the uncertainty on the mathematical function describing the reference spectral trends, as well as the uncertainties on the reference source's distance and \mbh, which were added in quadrature. 

\begin{figure}
 \includegraphics[width=\columnwidth]{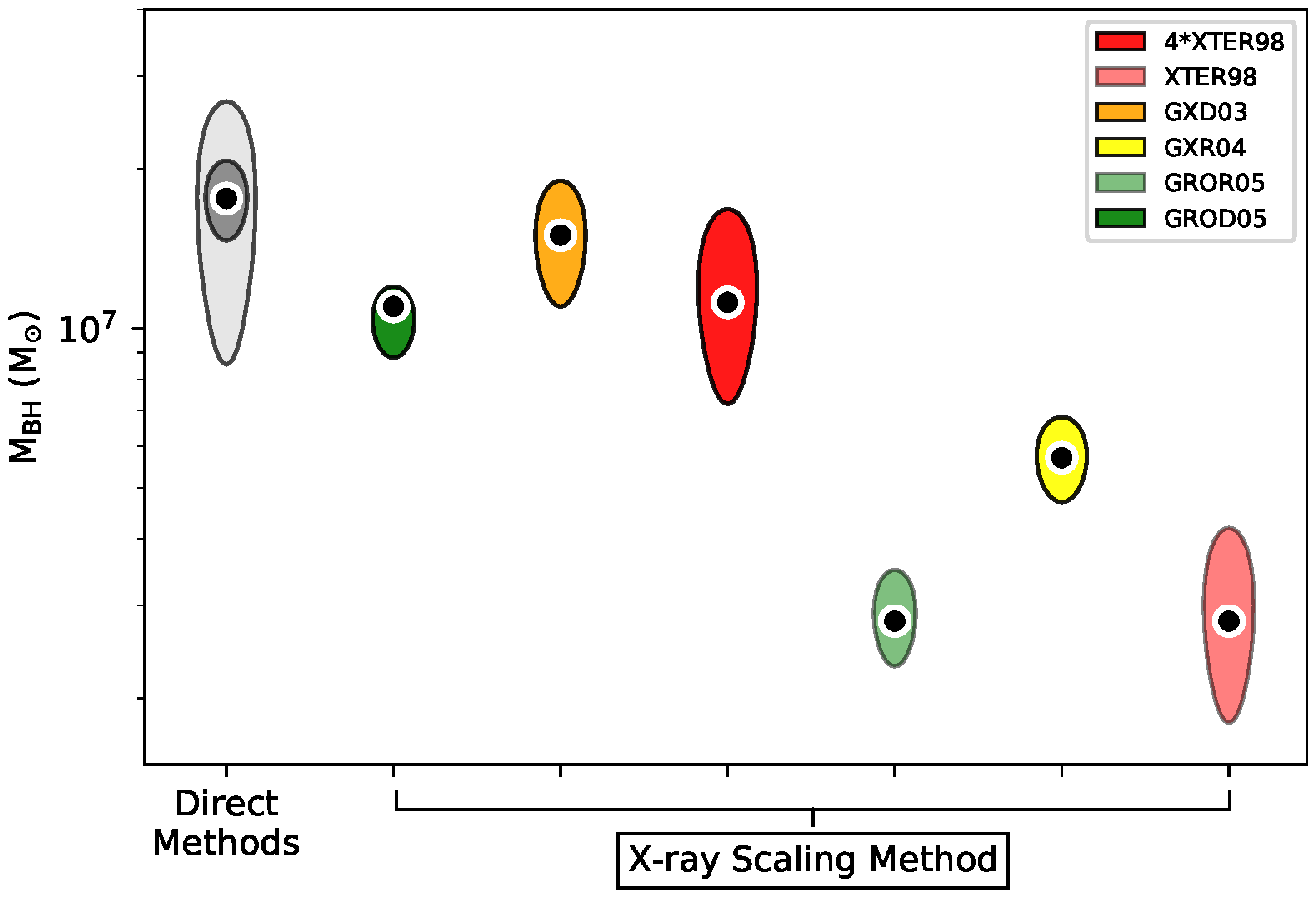}
 \caption{Comparison of the weighted mean \mbh\ value, obtained from the direct methods, with the estimates derived from the X-ray scaling method using different reference sources. The ellipses around the X-ray scaling values reflect the uncertainties on the AGN spectral parameters and the reference spectral trend, as well as the uncertainties on the \mbh\ and distance of the stellar references, as explained in the Appendix.}
 \label{fig:direct_vs_xray}
\end{figure}

The comparison of the average values derived using the scaling method with the different reference patterns and the weighted mean value obtained from independent dynamical methods is illustrated in Fig.~\ref{fig:direct_vs_xray} and reveals a good agreement (with the relative statistical uncertainties overlapping) of the values obtained with the GROD05, GXD03, and 4*XTER98 patterns.
\begin{table*}
	\centering
	\caption{\mbh\ estimates using the X-ray scaling method}
	\label{tab:Xscaling}
	\begin{tabular}{lccccccc}
		\hline
        \hline
		Reference source & \mbh\ Obs1 & \mbh\ Obs2 & \mbh\ Obs3 & \mbh\ Obs4 & \mbh\ average & \mbh\ min  & \mbh\ max\\
		& (\msol) & (\msol) & (\msol) & (\msol) & (\msol) & (\msol) & (\msol)\\
		(1) & \mcol{(2)} & \mcol{(3)} & \mcol{(4)} & \mcol{(5)} & \mcol{(6)} & \mcol{(7)} & \mcol{(8)}\\
		\hline
		\noalign{\smallskip}
		GROD05 & $1.0 \times 10^7$ & $1.1 \times 10^7$ & $1.1 \times 10^7$ & $9.7 \times 10^6$ & $1.1 \times 10^7$ & $8.8 \times 10^6$ & $1.2 \times 10^7$\\
		GXD03 & $1.4 \times 10^7$ & $1.6 \times 10^7$ & $1.6 \times 10^7$ & $1.4 \times 10^7$ & $1.5 \times 10^7$ & $1.1 \times 10^7$ & $1.9 \times 10^7$\\
		XTER98 & $2.6 \times 10^6$ & $3.1 \times 10^6$ & $2.9 \times 10^6$ & $2.5 \times 10^6$ & $2.8 \times 10^6$ & $1.8 \times 10^6$ & $4.2 \times 10^6$\\
		GROR05 & $2.7 \times 10^6$ & $2.9 \times 10^6$ & $3.0 \times 10^6$ & $2.6 \times 10^6$ & $2.8 \times 10^6$ & $2.3 \times 10^6$ & $3.5 \times 10^6$\\
		GXR04 & $5.4 \times 10^6$ & $6.6 \times 10^6$ & $5.9 \times 10^6$ & $5.1 \times 10^6$ & $5.7 \times 10^6$ & $4.7 \times 10^6$ & $6.8 \times 10^6$\\
		\hline
	\end{tabular}
	\begin{flushleft}
		Columns: 1 = reference source. 2--5 = \mbh\ measured for each observation. 5 = average of the \mbh\ values in columns 2--5. 6--7 = minimum and maximum mass estimates taking into account all the statistical errors and uncertainties described in Subsection \ref{sec:Xscaling}.
	\end{flushleft}
\end{table*}

\subsection{Fundamental plane of BH activity}
\label{sec:FP}
The fundamental plane of black hole activity (hereafter FP) was first presented by \citet{Merloni2003} for a heterogeneous sample of AGN and a few XRBs in the low/hard state, suggesting that astrophysical BHs at all scales follow a similar relationship involving the radio luminosity (universally associated with jet emission), the X-ray luminosity (generally associated with the accretion flow), and the \mbh. Although many different flavors of the FP have been proposed over the years focusing on different subsamples of AGN based either on the jet dominance or on specific ranges of the accretion rate (e.g., \citealt{Falcke2004,Yuan2005,Plotkin2012}), we will restrict our use of this method to determine the \mbh\ of \ngc\ to three versions of the FP: \\
\begin{enumerate}
    \item
    The original plane proposed by \citet{Merloni2003}, which, according to their equation (15), can be written as  
\begin{equation}
	\label{eq:Merloni}
    \log M_{BH} = 16.3 + \log D +1.28\left(\log F_R-0.60\log F_X \right)\pm 1.06
\end{equation}
where \mbh\ is in solar masses, $D$ is the distance in Mpc, $F_R$ the radio flux at 5 GHz, and $F_X$ the 2--10 keV absorption-corrected flux in $\mathrm{erg~s^{-1}~cm^{-2}}$.\\
    \item
     The FP restricted to relatively highly accreting AGN (including \ngc) and XRBs in the bright hard state proposed by \citet{Dong2014}. In this case, solving their equation (5) for \mbh, we obtain the following expression: 
\begin{equation}
	\label{eq:Dong}
    \log M_{BH} = 7.23\log L_X -4.55\log L_R - 131.68
\end{equation}
where \mbh\ is in solar masses, $L_R$ is the radio luminosity at 5 GHz, and $L_X$ the 2--10 keV intrinsic luminosity in $\mathrm{erg~s^{-1}}$. Note that in the original equation of \citet{Dong2014} the coefficient of $\log$\mbh\ has an uncertainty of nearly 100 per cent. This leads to unphysically large errors when solving the equation for \mbh\ and then propagating the errors.\\
     \item
     The FP based on the largest clean sample of AGN with direct \mbh\ measurements recently introduced by \citet{Gueltekin2019}, which explicitly utilizes the FP to constrain \mbh. Following their equation (21), we obtain: 
\begin{equation}
	\label{eq:Gueltekin}
    \mu=(0.55\pm0.22) + (1.09\pm0.10)R+\left(-0.59^{+0.16}_{-0.15}\right)X   
\end{equation}
where $\mu=\log(M_\mathrm{BH}/10^8\; \mathrm{M}_\odot)$, $R=\log(L_\mathrm{5~GHz}/10^{38}~\mathrm{erg\,s}^{-1})$, and $X=\log( L_\mathrm{2-10~keV}/10^{40}~\mathrm{erg\,s}^{-1})$.
\end{enumerate}

For the radio data, we used the 5 GHz VLA value provided by \citet{Fisher2021}, whereas for the X-ray we preferred the average value from the four \nustar\ observations described in Section~\ref{sec:Xanalysis} over the {\it Swift} XRT measurement obtained from an observation contemporaneous to the radio one. The reason is that the broader X-ray range covered by \nustar\ combined with a relatively large exposure (as opposed to the $ < 2$ ks of the XRT observation) makes it possible to better constrain the complex spectrum and hence the flux. In any case, a comparison of the \nustar\ average flux reveals that it is less than  
10 per cent higher than the XRT one. We can therefore conclude that the X-ray variability does not play any significant role in our calculation of \mbh\ with the FP.

The values of \mbh\ obtained with these different versions of the FP are reported in Table~\ref{tab:FP} with the minimum and maximum values derived from the uncertainties of the best-fitting parameters as well as from the measured scatter around the FP. Note that whereas \citet{Merloni2003} and \citet{Gueltekin2019} provide explicit equations of the FP solved for the \mbh, we had to solve equation (5) of \citet{Dong2014} for \mbh. For equation (5), we only considered the uncertainty on the numerical term and disregarded the uncertainties on the other two terms (the two containing logarithms). Otherwise, the uncertainty on the rewritten equation solved for \mbh\ would be too large. 

The results, illustrated in Fig.~\ref{fig:all_methods}, can be summarized as follows: the FP tends to underestimate the value of \mbh\ in \ngc, and the large scatter inherent to this method yields very large uncertainties, hampering the ability to tightly constrain \mbh\ with this method.

\begin{table*}
	\centering
	\caption{\mbh\ estimates using the fundamental plane of BH activity}
	\label{tab:FP}
	\begin{tabular}{lcccccc}
		\hline
        \hline
		Reference & Scatter & \mbh\ & \mbh\ min & \mbh\ max & \mbh\ min scatter & \mbh\ max scatter\\
		& (dex) & (\msol) & (\msol) & (\msol) & (\msol) & (\msol)\\
		(1) & \mcol{(2)} & \mcol{(3)} & \mcol{(4)} & \mcol{(5)} & \mcol{(6)} & \mcol{(7)}\\
		\hline
		\noalign{\smallskip}
		\citet{Merloni2003} & 0.88 & $9.2 \times 10^5$ & $8.0 \times 10^4$ & $1.1 \times 10^7$ & $1.2 \times 10^5$ & $7.0 \times 10^6$\\
		\citet{Dong2014} & 0.51 & $1.5 \times 10^7$ & $\dots$ &$\dots$ & $4.6 \times 10^6$ & $4.8 \times 10^7$\\
		\citet{Gueltekin2019} & 0.96 & $3.4 \times 10^6$ & $7.9 \times 10^5$ & $1.5 \times 10^7$ & $3.7 \times 10^5$ & $3.1 \times 10^7$\\
		\hline
	\end{tabular}
	\begin{flushleft}
		Columns: 1 = reference source. 2 = intrinsic scatter of \mbh. 3 = best \mbh\ estimate using the reference author's best-fitting equation. 4 = minimum mass estimate using the uncertainties on the equation's parameters. 5 = maximum mass estimate using the uncertainties on the equation's parameters. 6 = minimum mass estimate using the scatter. 7 = maximum mass estimate using the scatter. 
	\end{flushleft}
\end{table*}

\subsection{M--\texorpdfstring{\veldis}{sigma} relation}
\label{sec:Msigma}
Correlations between black hole masses and the properties of their host galaxies led some researchers to suspect that black holes and their host galaxies coevolve (for a survey, see \citet{Kormendy2013}). Correlations have now been found between \mbh\ and the mass, the luminosity, and the stellar velocity dispersion of the galaxy's bulge. In this section, we focus on the correlation between \mbh\ and the stellar velocity dispersion \veldis\ (see \citet{Gebhardt2000} and \citet{Ferrarese2000}) because that correlation is especially tight, which is important because it is easier to measure \veldis\ than it is to measure \mbh\ directly by other methods. In this section only, we use \veldis\ for the velocity dispersion to distinguish it from the usual standard deviation $\sigma$; in the rest of the paper, there is no confusion between the symbols. In this section, we estimate the black hole mass of NGC 4151 using models for the M--\veldis\ relationship found in the four papers we describe in the rest of this section.

All four papers use a best-fitting equation of the form:
\begin{equation}
	\label{eq:McConnell}
    \log \left( \frac{M_{\mathrm{BH}}}{\mathrm{M}_\odot} \right) = \alpha + \beta \log \left( \frac{\sigma_\star}{200\, \mathrm{km\, s}^{-1}} \right)
\end{equation}
The authors, however, use different samples and therefore get different fitted values of $\alpha$, $\beta$, and intrinsic scatter. \citet{Kormendy2013} use a scaling factor of $10^9$ in the denominator of the left-hand side; the other three papers do not. We use all four of these fits to estimate \mbh. Throughout this paper, we let $\sigma_\star = 91.8 \pm 9.9$ km\,s$^{-1}$, the value found in the Hyperleda database for \ngc.
 
\citet{McConnell2013} used a more recent sample of 72 black holes and their host galaxies to revise the M--\veldis\ relation. They found slightly different fits for late- and early-type galaxies. Since \ngc\ is an Sb spiral galaxy, we use their best-fitting model for late-type galaxies, which gives $\alpha = 8.07 \pm 0.21$ and $\beta = 5.06 \pm 1.16$ (see their table 2). Plugging these values into equation~(\ref{eq:McConnell}) gives the values of \mbh, \mbh\ min, and \mbh\ max shown in Table~\ref{tab:m-sigma}. For the minimum and maximum values, we simply add uncertainties. However, because the logarithm in equation~(\ref{eq:McConnell}) is negative for our value of $\sigma_\star$, the minimum value of \mbh\ occurs at ($\alpha-\sigma_\alpha$, $\beta+\sigma_\beta$, $\sigma_\star-\sigma_{\sigma_\star}$). By a similar argument, the maximum occurs at ($\alpha+\sigma_\alpha$, $\beta-\sigma_\beta$, $\sigma_\star+\sigma_{\sigma_\star}$). Here, the unstarred sigmas are the usual standard deviations, the uncertainties on the parameters $\alpha$, $\beta$, and \veldis\ in the equation. That leaves scatter to consider. \citet{McConnell2013} calculated an intrinsic scatter in \mbh\ of 0.46 dex for their model. If we apply this directly to $\log M_{BH}$, we get the values of `\mbh\ min scatter' and `\mbh\ max scatter' in Table~\ref{tab:m-sigma}. Here we are ignoring the errors on $\alpha$ and $\beta$ and using only scatter to compute the minimum and maximum values of \mbh. 

\citet{Woo2013} looked at the M--\veldis\ relation from another point of view. They asked if the relation is the same for quiescent galaxies and active ones and found that, after accounting for selection effects, it is the same for both. Woo's team distinguished many cases of galaxies and calculated parameters for equation~(\ref{eq:McConnell}) for each of them. Here we use the following values from the `reverberation-mapped AGN' line of their table 3: $\alpha = 7.31 \pm 0.15, \beta = 3.46 \pm 0.61$, and scatter = 0.41 dex. Plugging these values into equation~(\ref{eq:McConnell}) just as we did for the McConnell and Ma model, we calculated \mbh, \mbh\ min and max, and \mbh\ min and max scatter. Our results are given in Table~\ref{tab:m-sigma}.

\citet{Kormendy2013} relooked at the M--\veldis\ correlation with the idea of getting an even better correlation by throwing out several classes of outliers. They showed that galaxies with pseudobulges do not have the same tight correlations seen in ellipticals and in spirals with classical bulges, and therefore galaxies with pseudobulges were excluded. Kormendy's team also noted that black hole masses derived from emission-line rotation curves are too low unless broad line widths are also accounted for, so they excluded those as well. They also excluded mergers in progress, which have especially low masses, and two outliers with very high-mass black holes, which they called `BH monsters'. After excluding all these outliers, they got the following best fit: $\alpha = -0.510 \pm 0.049$, $\beta = 4.377 \pm 0.290$, and scatter = 0.29 dex (see their equation 3). Plugging these values into equation~(\ref{eq:McConnell}) just as we did before, we calculated \mbh, \mbh\ min and max, and \mbh\ min and max scatter. The results are given in Table~\ref{tab:m-sigma}.

The fourth and final paper we used to estimate \mbh\ for \ngc\ using the M--\veldis\ relation is \citet{Shankar2016}. Shankar's team claimed that difficulties in resolving the sphere of influence of the BH caused selection biases in previous samples of BHs, which favored higher-than-usual \veldis\ values. This selection effect means that earlier models of the M--\veldis\ relation were biased. When this is corrected for, the authors give the following best fit for what they claim is the intrinsic M--\veldis\ relation (see their equation 7 of their model 1): $\alpha = 7.8$ and $\beta = 5.7$. The authors assume a Gaussian scatter of 0.25 dex. Plugging these values into equation~(\ref{eq:McConnell}), we calculated the same five values as before, which are given in Table~\ref{tab:m-sigma}. 

The visual comparison of the \mbh\ values obtained by the different prescriptions of the M--\veldis\ relation and the value constrained dynamically is shown in Fig.~\ref{fig:all_methods}, revealing that only the value derived from the version of \citet{Kormendy2013} appears to be consistent with the dynamical values, whereas the others underestimate the \mbh.

\begin{table*}
	\centering
	\caption{\mbh\ estimates using the M-sigma relation}
	\label{tab:m-sigma}
	\begin{tabular}{lcccccc}
		\hline
        \hline
		Reference & Scatter & \mbh\ & \mbh\ min & \mbh\ max & \mbh\ min scatter & \mbh\ max scatter\\
		& (dex) & (\msol) & (\msol) & (\msol) & (\msol) & (\msol)\\
		(1) & \mcol{(2)} & \mcol{(3)} & \mcol{(4)} & \mcol{(5)} & \mcol{(6)} & \mcol{(7)}\\
		\hline
		\noalign{\smallskip}
		\citet{Woo2013} & 0.41 & $1.4 \times 10^6$ & $3.8 \times 10^5$ & $4.2 \times 10^6$ & $5.4 \times 10^5$ & $3.6 \times 10^6$\\
		\citet{McConnell2013} & 0.46 & $2.3 \times 10^6$ & $2.8 \times 10^5$ & $1.4 \times 10^7$ & $7.9 \times 10^5$ & $6.6 \times 10^6$\\
		\citet{Kormendy2013} & 0.29 & $1.0 \times 10^7$ & $4.3 \times 10^6$ & $2.2 \times 10^7$ & $5.3 \times 10^6$ & $2.0 \times 10^7$\\
		\cite{Shankar2016} & 0.25 & $7.5 \times 10^5$ & \ldots & \ldots & $4.2 \times 10^5$ & $1.3 \times 10^6$\\
		\hline
	\end{tabular}
	\begin{flushleft}
		Columns: 1 = reference source. 2 = intrinsic scatter of \mbh. 3 = best \mbh\ estimate using the reference author's best-fitting equation. 4 = minimum mass estimate using the uncertainties on the equation's parameters. 5 = maximum mass estimate using the uncertainties on the equation's parameters. 6 = minimum mass estimate using the scatter. 7 = maximum mass estimate using the scatter. \mbh\ min and \mbh\ max are blank for the last row because Shankar does not assign errors to the values of $\alpha$ and $\beta$. See Section \ref{sec:Msigma} for a fuller explanation of this table.
	\end{flushleft}
\end{table*}

\section{Discussion and Conclusions}
\label{sec:discussion}
In this work we have used \ngc, one of the few sources for which \mbh\ has been measured with different independent direct methods, as a laboratory to test three indirect methods that can be used to constrain \mbh\ in AGN that are heavily obscured. Before making a systematic comparison between these indirect methods, we briefly discuss the results from the direct dynamical methods. 

\begin{figure*}
 \includegraphics
 {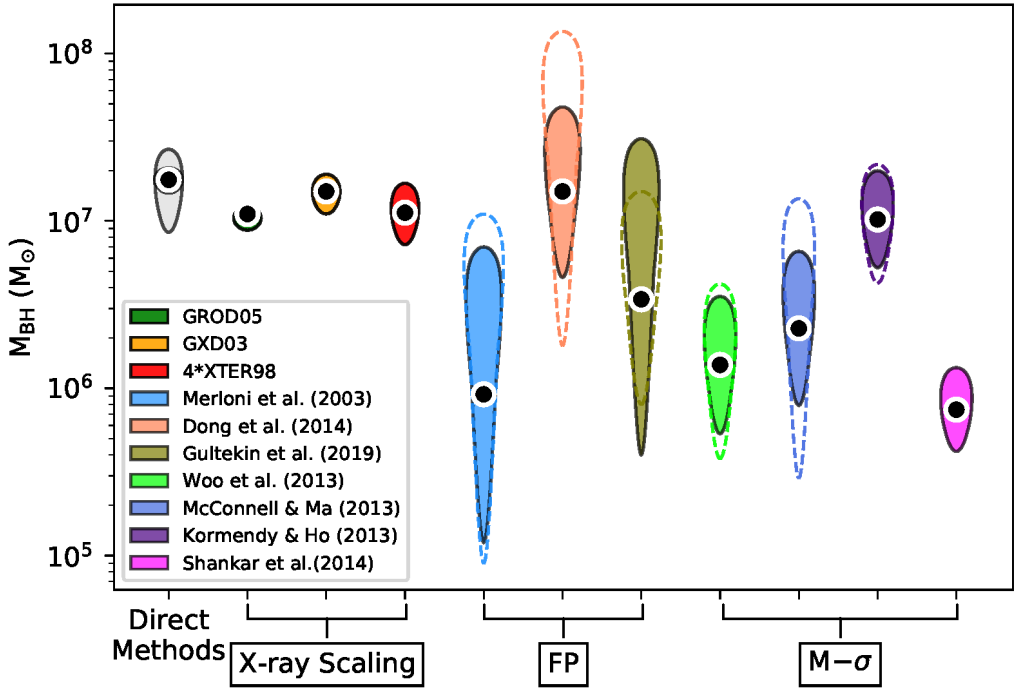}
 \caption{Comparison between the weighted mean ${\langle M_{\mathrm BH}\rangle}$ obtained with all the available independent direct methods and the values obtained with the X-ray scaling method, with the fundamental plane of black hole activity, and with different incarnations of the M--$\sigma$ correlation. The uncertainties of the FP and M--$\sigma$ values, illustrated by the color-filled ellipses, represent the scattering of these correlations; the dashed ellipses reflect the uncertainties on the equation parameters. For the \citet{Dong2014} estimate the dashed ellipse represents the scattering derived by \citet{Gueltekin2019}, which we deemed more appropriate to describe the actual uncertainty of the FP method (see Discussion).}
 \label{fig:all_methods}
\end{figure*}

Interestingly enough, despite the very different assumptions of each method, the \mbh\ measurements from the different dynamical methods are consistent with each other within their respective statistical uncertainties, which range from 25 per cent \citep{Bentz2022} to 90 per cent \citep{Zoghbi2019}. Despite the overall agreement, these \mbh\ values vary by a factor of $\sim$3. Therefore, to perform a systematic comparison with the values obtained from the three indirect methods, we have computed the weighted mean, ${\langle M_{\mathrm BH}\rangle}=1.76\times 10^7~ (\sigma=3.05\times10^6)~\mathrm{M_\odot}$, using one measurement for each independent direct method. For methods with more than one measurement, we used the most recent one, under the assumption that they are characterized by higher quality data and/or improvements in the modeling techniques compared to the previous measurements. For the H$_\beta$ reverberation method, we utilized both the result from \citet{Derosa2018} and the one from \citet{Bentz2022}, because the techniques used in these two papers are different: the former uses an average virial factor to derive \mbh, whereas the latter derives \mbh\ from the direct modeling of the BLR kinematics. For completeness, we report also the weighted mean, obtained by including the older stellar dynamics value from \citet{Onken2014} and the first H$_\beta$ reverberation value from \citet{Bentz2006}. In any case, the value obtained in this way, ${\langle M_{\mathrm BH}\rangle}=1.96\times 10^7~ (\sigma=2.91\times10^6)~\mathrm{M_\odot}$, is fairly similar to the one derived without these two older measurements, and hence our conclusions are not affected by the exclusion of these two direct measurements.

A visual comparison between the weighted mean, obtained using the most recent measurement for each of the independent direct methods, and the values derived with the three indirect methods tested here is illustrated in Fig.~\ref{fig:all_methods}. This figure reveals that the values obtained with the X-ray scaling method using the GROD05 and GXD03 reference patterns are
fully consistent with ${\langle M_{\mathrm BH}\rangle}$ and are characterized by small statistical uncertainties. For completeness, we also show the value obtained with the XTER98 pattern, which must be calibrated by a multiplicative factor of 4 to be consistent with the other two measurements from this method and hence with ${\langle M_{\mathrm BH}\rangle}$. As explained in detail in the Appendix, it is important to calibrate this additional reference pattern, because it makes it possible to constrain the \mbh\ of AGN with $\Gamma$ ranging between 1.4 and $\sim$3, whereas the two most reliable patterns, GROD05 and GXD03, can only be used for AGN with X-ray photon indices between $\sim$1.5 and 2.

Fig.~\ref{fig:all_methods} also shows that the values obtained with different prescriptions of the fundamental plane of BH activity tend to underestimate \mbh\ compared to the mean from direct methods and to have large statistical uncertainties, which reflect the relatively large scatter observed in the FP.  We note that of the three incarnations of the FP tested in this work, the one from \citet{Dong2014} yields the closest estimate, which is perhaps not surprising, since that FP is limited to radiatively efficient sources and \ngc\ is part of the AGN sample used to build that FP. What is perhaps a bit surprising is that the scatter reported by \citet{Dong2014}, 0.51 dex, using a sample of AGN with archival non-contemporaneous X-ray and radio data and with \mbh\ obtained with heterogeneous secondary methods, is substantially smaller than the one measured by \citet{Gueltekin2019} (0.96 dex), using a carefully selected sample of AGN with \mbh\ measured via direct primary methods, with contemporaneous X-ray and high-resolution radio data. We, therefore, consider $\sim$1 dex as a more realistic uncertainty for this method, and confirm that this technique can only provide a crude estimate of the \mbh\ and is better suited to discriminate between stellar and supermassive BHs. Finally, it is worth noting that one of the prescriptions for the correct application of the FP is the requirement to use high angular resolution data to probe the core emission of the source; however, recent studies show that, when milliarcsecond resolution VLBA data are utilized instead of the lower resolution VLA data, the mostly type 2 AGN do not follow the FP any more \citep{Fisher2021,Shuvo2022}.

From Fig.~\ref{fig:all_methods} it is also clear that the values obtained from the M--$\sigma$ correlation, which have a considerably smaller uncertainty compared to the FP ones, tend to underestimate \mbh\ by almost an order of magnitude with the notable exception of the value yielded by the \citet{Kormendy2013} prescription, which is formally consistent with ${\langle M_{\mathrm BH}\rangle}$. Based on the comparison with the dynamical value of \ngc, we might conclude that the M--$\sigma$ correlation of \citet{Kormendy2013} offers the quickest accurate method to determine \mbh. However, our original goal was to assess the reliability of indirect methods to constrain \mbh\ in heavily obscured AGN, where direct methods (or indirect methods based on BLR spectral measurements) cannot be applied, and there is suggestive evidence that the M--$\sigma$ correlation may significantly overestimate the \mbh\ in type 2 AGN (see, e.g., \citealt{Greene2016,Ricci2017}). Therefore, the systematic application of the M--$\sigma$ correlation of \citet{Kormendy2013} (which yields values larger than the other correlations considered here) to type 2 AGN should be considered with caution.

A more quantitative comparison of the \mbh\ values obtained from the three indirect methods and the weighted mean from the multiple direct measurements is provided by the ratio 
$M_{\mathrm{BH,ratio}}={\langle M_{\mathrm{BH,direct}}\rangle}/ M_{\mathrm{BH,indirect}}$ and by the per cent statistical uncertainty of each \mbh\ measurement. The numerical values for these quantities are reported in Table~\ref{table:comparison}, where we use boldface to indicate estimates deemed {\it accurate}, based on the criterion $M_{\mathrm{BH,ratio}}<3$, and {\it precise}, based on the criterion per cent error < 100 per cent. According to these two criteria, only the values from the X-ray scaling method and the one from the M--$\sigma$ correlation of \citet{Kormendy2013} are both accurate and precise, whereas the fundamental plane of BH activity of \citet{Dong2014} provides an accurate estimate of the \mbh\ but not a precise one. 

It is reassuring that three very different indirect methods yield values of \mbh\ consistent with the ones obtained from direct dynamical methods. Indeed, when dealing with heavily obscured AGN, which usually have no direct \mbh\ measurements, the agreement between independent indirect methods may be the only viable approach to obtain a reliable estimate.  It is important to note that these three indirect methods have very different assumptions and are valid in different regimes of \mbh\ and $\dot m$. 

For example, the X-ray scaling method, which was developed for stellar-mass BHs \citep{Shaposhnikov2009} and later extended to supermassive BHs in both type 1 and type 2 AGN \citep{Gliozzi2011,Gliozzi2021}, has no limits in \mbh. However, since this method relies on the softer-when-brighter spectral trend, it can only be applied to BH systems accreting at moderate or high rate, with a radiatively efficient accretion disk coupled with a hot corona. For very low-accreting sources ($\dot m \ll 0.01$), the X-rays are produced by a radiatively inefficient accretion flow with the possible contribution from the base of a jet and follow an opposite spectral trend; therefore, the scaling method cannot be extended to this low-accretion regime, as explicitly shown by \citet{Jang2014}. Additionally, when dealing with obscured AGN, a broadband good-quality X-ray spectrum is necessary to derive realistic estimates of the primary X-ray emission and thus of \mbh\ (this limitation applies as well to the FP method).

Also, the FP method can in principle be extended to any \mbh\ since by construction it encompasses XRBs and supermassive BHs. In terms of $\dot m$, different versions of the FP have been developed with specific BH samples: for radiatively efficient sources, the FP proposed by \citet{Dong2014} appears to be the most appropriate, whereas for low-accreting sources, the FP presented by \citet{Gueltekin2019} yields the most reliable estimates. The main limitations of this method are the requirement of high-resolution radio data preferably contemporaneous with the X-ray observations and the inherently large scatter associated with the FP that yields only crude estimates of \mbh.

The M--$\sigma$ correlation by construction can only be applied to supermassive BHs at the centers of galaxies. Additionally, there is evidence that $\sigma$ saturates at high \mbh\ values, and it is unclear if the correlation extends to the very low-mass BH regime, where the host galaxies have no classical bulges \citep{Kormendy2013}. Besides these limitations in the \mbh\ range, it is unclear if the M--$\sigma$ correlation can be applied to a wide range of $\dot m$. Having been developed for (nearly) quiescent galaxies, the M--$\sigma$ correlation is certainly appropriate for the low-accreting regime, and its successful application to type 1 AGN suggests that it can be extended to higher accretion rate values \citep{Woo2013}. However, in obscured AGN with \mbh\ dynamically determined via maser or via IR measurements of the BLR, the M--$\sigma$ correlation appears to systematically overestimate the \mbh\ \citep{Greene2016,Ricci2017}, casting some doubts on its reliability for type 2 AGN.

To conclude, we summarize the main results of this work, where we used the nearby X-ray bright Seyfert galaxy \ngc\ as a laboratory to compare and assess the reliability of three indirect methods (the X-ray scaling method, the fundamental plane of black hole activity, and the M--$\sigma$ correlation) that can be utilized to estimate \mbh\ in extragalactic sources where direct methods cannot be applied, as in the case of heavily obscured AGN.\\
\begin{itemize}
    \item All three methods -- specifically, the X-ray scaling method with the reference patterns GROD05 and GXD03, the M--$\sigma$ correlation of \citet{Kormendy2013}, and the FP of radiatively efficient BHs of \citet{Dong2014} -- yield \mbh\ values formally consistent with those obtained from direct dynamical methods and hence can be considered accurate. However, only the first two have relatively small statistical uncertainties and can be deemed precise.\\
    \item The agreement of the \mbh\ dynamical estimates with the values obtained using three independent indirect methods, which are based on very different assumptions, lends support to the possibility of obtaining reliable \mbh\ values in AGN that are not accessible to direct dynamical methods.\\
    \item Each indirect method has different limitations in terms of \mbh\ and $\dot m$ ranges. For instance, the X-ray scaling method has no upper or lower limits in \mbh\ but cannot be applied to low-accreting systems ($\dot m \ll 0.01$). On the other hand, the M--$\sigma$ correlation can be applied to any accretion regime but has both upper and lower limits on \mbh, because of the saturation of $\sigma$ at high masses and the lack of classical bulges at low masses. Finally, the FP has no limitations on \mbh, and using the appropriate version can in principle be applied to both high- and low-accreting systems; however, the intrinsic uncertainty of this method makes it useful only for crude \mbh\ estimates.
\end{itemize}

We conclude that the best approach to determine \mbh\ in systems not accessible to direct methods is to apply independently the three methods (with different prescriptions for the M--$\sigma$ correlation and for the FP) accounting for their respective limitations and range of applicability. In particular, when estimating \mbh\ in heavily obscured AGN, one must bear in mind that the M--$\sigma$ correlation tends to overestimate \mbh\ \citep{Greene2016,Ricci2017} and the prescription from \citet{Kormendy2013}, which yields the largest values among the M--$\sigma$ correlations, exacerbates this problem.

\begin{table}
	\caption{\mbh\ comparison}		
	\begin{center}
	\begin{tabular}{llrr} 
			\hline
			\hline       
			Method & transition/reference & $M_{\mathrm{BH,ratio}}$ & \% error on \mbh\ \\
			(1) & (2) & (3) & (4) \\
			\hline
			X-ray scaling & GROD05 &  {\bf 1.7} & {\bf 17\%} \\
			X-ray scaling & GXD03 &  {\bf 1.2} & {\bf 25\%} \\
			X-ray scaling & 4*XTER98 &  {\bf 1.6} & {\bf 44\%} \\
			FP & \citet{Merloni2003} & 19.2 & 373\% \\
			FP & \citet{Dong2014} & {\bf 1.2} & 146\% \\
			FP & \citet{Gueltekin2019} & 5.1 & 450\% \\
			M-$\sigma$ & \citet{Woo2013} & 12.8 & 109\% \\
			M-$\sigma$ & \citet{McConnell2013} & 7.7 & 127\% \\
			M-$\sigma$ & \citet{Kormendy2013} & {\bf 1.7} & {\bf 72\%} \\
			M-$\sigma$ & \citet{Shankar2016} & 23.6 & 61\% \\
			\hline
		\end{tabular}	
	\begin{flushleft}
		Columns: 1 = method. 2 = transition (for the X-ray scaling method) or reference (for the FP and M--$\sigma$ methods). 3 = ratio of ${\langle M_{\mathrm{BH,direct}}\rangle}$ to $M_{\mathrm{BH,indirect}}$. Boldface in column 3 indicates estimates deemed accurate, based on the criterion $M_{\mathrm{BH,ratio}}<3$. 4 = average percentage error computed by each method. Boldface in column 4 indicates estimates deemed precise, based on the criterion \% error < 100\%.
	\end{flushleft}
	\end{center}
	\label{table:comparison}
\end{table}

\section*{Acknowledgements}

 George Mason University OSCAR Undergraduate Research Scholarship Program funding was used in the X-ray spectral analysis portion of this work. We thank the anonymous referee for an especially thorough and constructive review, which improved the clarity of the paper. 

\section*{Data Availability}

The data underlying this article are available in the High Energy Astrophysics Science Archive Research Center (HEASARC) archive at https://heasarc.gsfc.nasa.gov/docs/archive.html.


\bibliographystyle{mnras}
\typeout{}
\bibliography{astro}




\appendix

\section{X-ray scaling method: a quick guide}
We first derive the equation used to determine \mbh\ with the X-ray scaling method and explain the underlying assumptions. Then, we provide the essential information about the stellar reference sources used in this work. Finally, we list the individual steps needed to derive \mbh\ starting from the X-ray spectral results.\\

\noindent {\bf Basic equations and assumptions.}
The first (and universally accepted) assumption is that the bulk of the X-ray emission is produced by Comptonization in the corona surrounding the accretion flow. Therefore, the first step is to fit the X-ray spectrum with the \textsc{bmc} model, a simple Comptonization model that can be used to describe both the bulk motion and the thermal Comptonization \citep{Titarchuk1997}. This model is characterized by four parameters, the normalization \nbmc\ (defined as the accretion luminosity in units of $10^{39}$ erg s$^{-1}$ divided by the distance squared in units of 10 kpc), the spectral index $\alpha$ (related to the photon index by the relationship $\Gamma=\alpha+1$), the temperature of the seed photons $kT$ (which for AGN is typically restricted to a narrow range between 0.1 and 0.3 keV), and $\log A$, where $A$ is related to the fraction of scattered seed photons $f$ by the relationship $f=A/(A + 1)$. The last parameter, $\log A$, is generally poorly constrained but has very little impact on the determination of \mbh. 

In this method the BH mass is obtained by scaling up the normalization of the stellar reference source $N_{\mathrm{BMC,ref}}$ to the corresponding normalization of the AGN $N_{\mathrm{BMC,AGN}}$. In other words, the starting equation is the ratio between the \textsc{bmc} normalizations:
\[
\left(\frac{N_{\mathrm{BMC,AGN}}}{N_{\mathrm{BMC,ref}}}\right) =
 \left(\frac{L_{\mathrm{AGN}}}{L_{\mathrm{ref}}}\right) \times \left(\frac{d_{\mathrm{ref}}^2}{d_{\mathrm{AGN}}^2}\right)
\]

Since the accretion luminosity depends on the radiative efficiency, accretion rate, and BH mass, the above equation can be rewritten as:
\[
\left(\frac{N_{\mathrm{BMC,AGN}}}{N_{\mathrm{BMC,ref}}}\right) = \left(\frac{(\eta \times \dot{m} \times M_{\mathrm{BH}})_{\mathrm{AGN}}}{(\eta \times \dot{m} \times M_{\mathrm{BH}})_{\mathrm{ref}}}\right) \times \left(\frac{d_{\mathrm{ref}}^2}{d_{\mathrm{AGN}}^2}\right)
\]
where $\eta$ is the radiative efficiency and $\dot{m}$ the accretion rate in Eddington units for the AGN and the reference source, respectively.

The second fundamental assumption of this method is that the photon index is a reliable indicator of the accretion state of BH systems at all scales. In other words, two BH systems with the same $\Gamma$ are considered to be in the same accretion state, and hence to have the same accretion rate (in Eddington units) and radiative efficiency. Note that this assumption (and hence the applicability of the X-ray scaling method) is limited to BHs accreting at or above the 1 per cent threshold in Eddington units. Below this threshold the  softer-when-brighter trend vanishes, and at even lower values of $\dot{m}$ an inverse correlation is observed (see e.g, \citealt{Gu2009,Constantin2009,Jang2014}). With this second assumption, solving the above equation for \mbh\, we obtain the final equation:
\begin{equation}\label{eq_Xscal}
M_{\mathrm{BH,AGN}}=M_{\mathrm{BH,ref}}  \times \left(\frac{N_{\mathrm{BMC,AGN}}}{N_{\mathrm{BMC,ref}}}\right)  \times \left(\frac{d_{\mathrm{AGN}}^2}{d_{\mathrm{ref}}^2}\right)
\end{equation}
which implies that the AGN \mbh\ is simply obtained by multiplying the dynamical mass of the XRB reference times the ratio of their respective distances squared, times the ratio of their \nbmc\ values.\\

\noindent {\bf Reference sources.} The fundamental properties of the three reference sources are reported in Table~\ref{table:tabA1}, whereas the functional form of the spectral trends of the reference sources, necessary to compute the \nbmc\ ratio, is described by the following equation:
\begin{equation}\label{eqjang}
	N_\textrm{BMC,ref}(\Gamma) = N_\textrm{tr} \times \left\lbrace 1 - \ln \left[ \exp \left( \frac{A - \Gamma}{B} \right) - 1 \right] \right\rbrace ^{(1/\beta)}
\end{equation}
where $\Gamma$ is the photon index of the target AGN as determined by the spectral fit, and $A$, $B$, $N_\mathrm{tr}$, and $\beta$ are given in Table~\ref{table:tabA2}.  

\begin{figure}
 \includegraphics[width=\columnwidth]{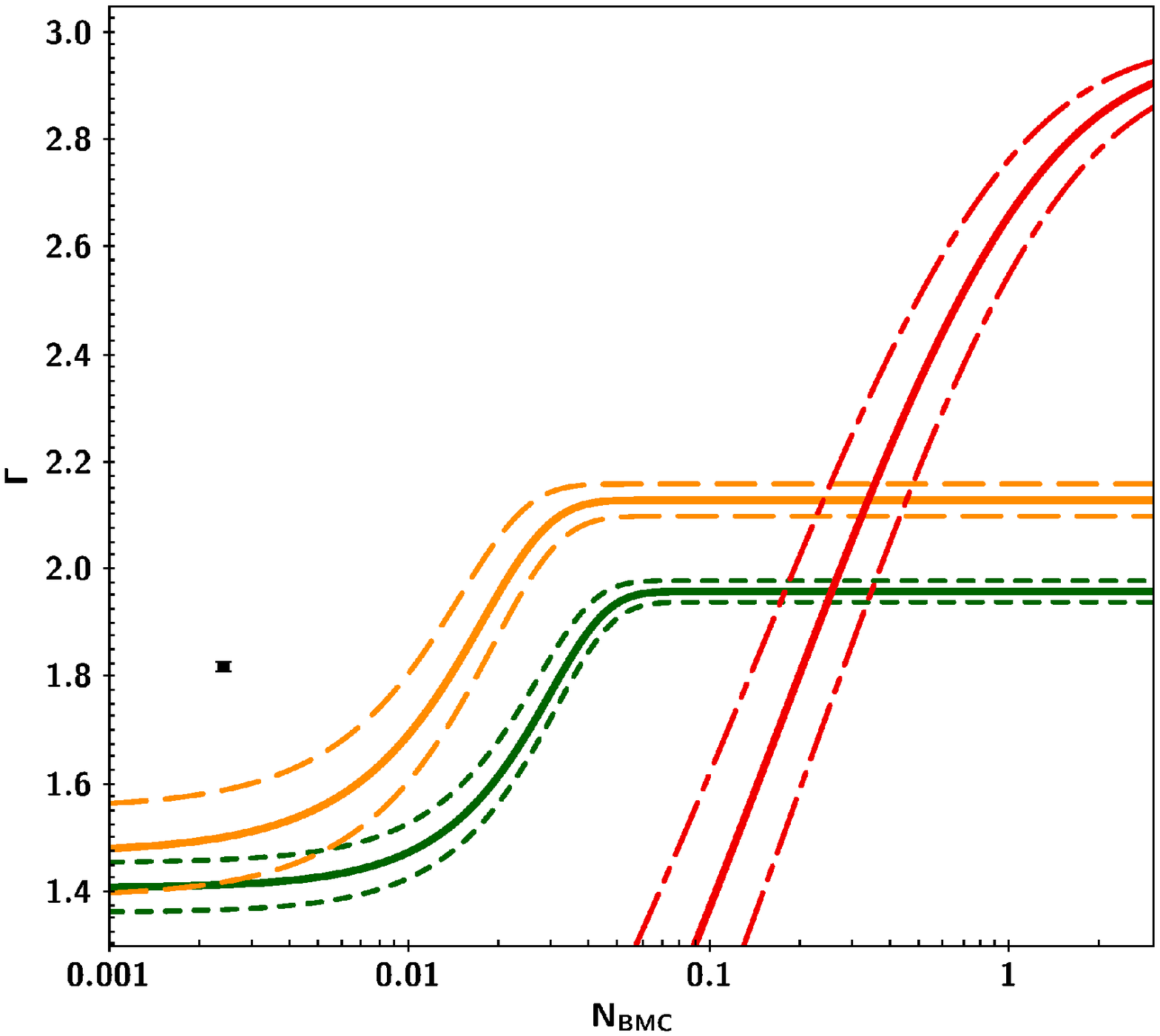}
 \caption{\nbmc\ -- $\Gamma$ diagram with the three most reliable reference sources \gro\ during the 2005 decay (in dark green), \gx\ during the 2003 decay (orange), and \xte\ during the 1998 rise (red). The continuous lines indicate the best fit spectral trends, whereas the short-dashed, long-dashed, and short-long-dashed lines indicate the 1$\sigma$ uncertainties for \gro, \gx, and \xte, respectively. Based on the best fit of the \nustar\ spectrum in Obs3 (the observation with the largest flux), the location of \ngc\ in this diagram is represented by the black point (the error bars are comparable with the symbol size). }
 \label{fig:Xreferences}
\end{figure}

These trends are illustrated in Fig.~\ref{fig:Xreferences} for GROD05 (dark green line), GXD03 (orange), and XTER98 (red). From the figure it is clear that the first two trends are similar and hence yield similar values of the AGN \mbh, whereas the third one gives comparatively smaller values. However, the XTER98 spectral trend has a much larger range in $\Gamma$ and hence allows the application of this method to basically any AGN (as opposed to GROD05 and GXD03, which can only be applied to AGN with $\Gamma\sim$ 1.5--2). For this reason it is important to calibrate this third trend: a comparison with the \mbh\ values obtained from GROD05 and GXD03 implies that the calibration factor should be around 4, which is similar to the calibration value of 3 derived in \citet{Gliozzi2011} using a reverberation mapping sample of AGN.\\

\noindent {\bf Computing \mbh\ in three steps.} First, it is necessary to fit the primary X-ray spectral component with the \textsc{bmc} model, after properly accounting for absorption and reflection. For example, using the spectral data of Obs3 of \ngc\ yields $\Gamma=1.83\pm0.01$ and $N_{\rm BMC}=(2.40\pm0.01)\times 10^{-3}$.

Second, one should derive the \nbmc\ value for the reference source at the appropriate value of $\Gamma$. This can be done quickly by visually inspecting Fig.~\ref{fig:Xreferences} and finding the approximate values of \nbmc\ corresponding to $\Gamma=1.83$ for the different reference sources ($\sim0.03$ for GROD05, $\sim0.015$ for GXD03, and $\sim0.2$ for XTER98). Alternatively, one can plug the value of $\Gamma=1.83$ into equation~(\ref{eqjang}) and obtain the exact values. For the sake of simplicity, we also include the tabulated \nbmc\ values of these three reference sources in Table~\ref{tab:nbmc}.

\begin{figure}
 \includegraphics[width=\columnwidth]{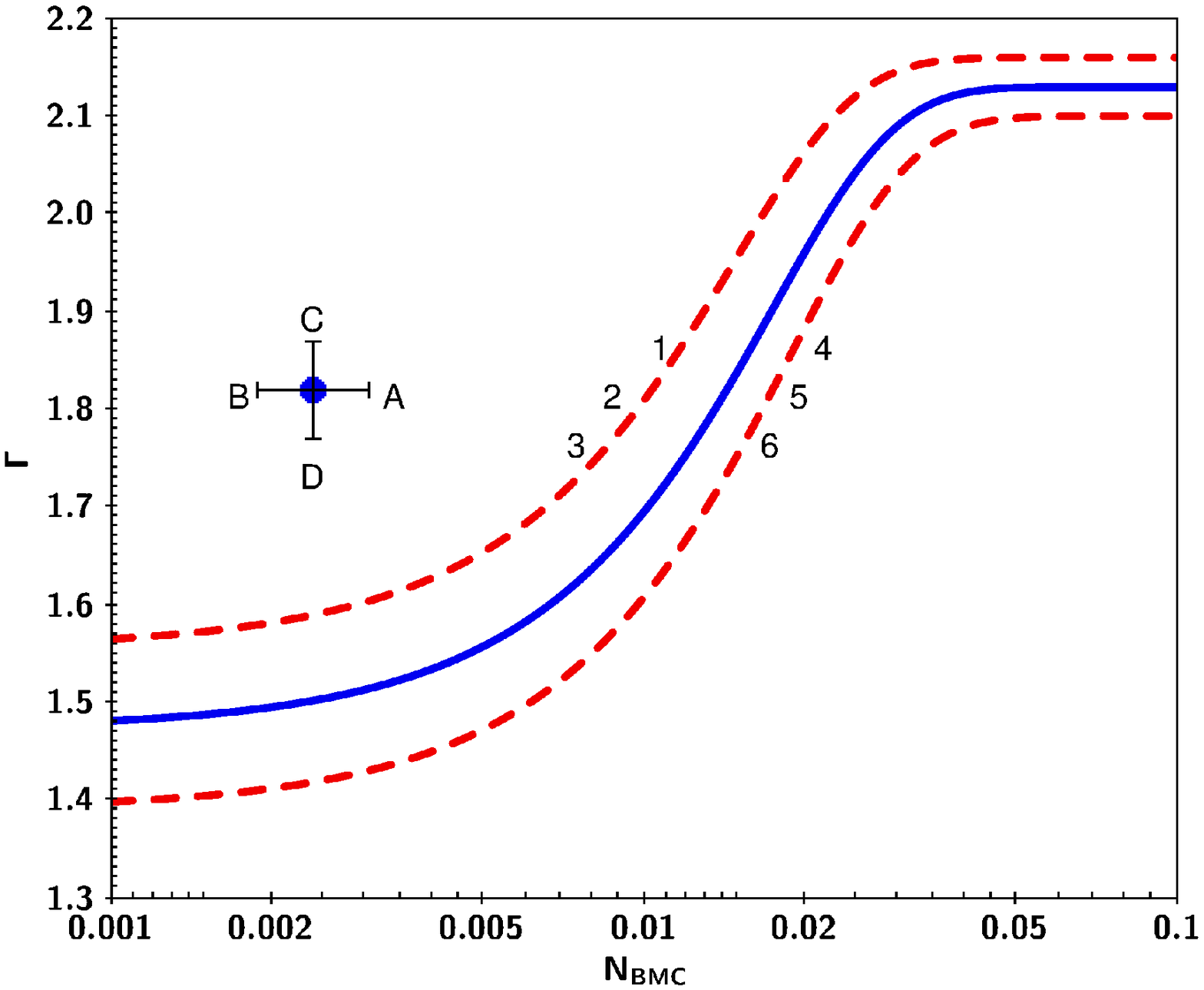}
 \caption{ \nbmc\ -- $\Gamma$ diagram used to illustrate how the uncertainties on \mbh\ are computed taking into account the uncertainties on the spectral parameters $\Gamma$ and \nbmc, as well as the uncertainties on the spectral trends of the reference sources. For illustration purposes the statistical uncertainties on $\Gamma$ and \nbmc\ have been increased by a factor of 5 and 10, respectively. The figure includes the best-fitting trend of the reference source \gx\ during the 2003 decay, described by the continuous blue line and its 1$\sigma$ envelope shown by the dashed red lines. For each of the points A, B, C, and D of the AGN, the separation from the corresponding points on the 1$\sigma$ envelope of the reference source trend (described by the points 1, 2, 3, 4, 5, 6) is computed. This is the ratio between the AGN and reference source \nbmc\ used in  equation~(\ref{eq_Xscal}) to compute \mbh. From all these \mbh\ values the minimum and the maximum are used to compute the lower and upper uncertainty of the X-ray scaling method. For completeness, the final uncertainty also includes the fractional uncertainties on the reference \mbh\ and distance (given in Table~\ref{table:tabA1}), which are added in quadrature to the fractional error associated with the spectral uncertainty described above.}
 \label{fig:Xrerror}
\end{figure}

Finally, from Table~\ref{table:tabA1} one must insert into equation~(\ref{eq_Xscal}) all the numerical values including the appropriate reference's \mbh\ and distance. For GROD05, plugging in \mbh$_{\textrm{,ref}}=6.3~\mathrm{M}_\odot$, $d_{\mathrm{ref}} = 3.2$ kpc, and \nbmc$_{\textrm{,ref}}=0.034$, along with $d_{\mathrm{AGN}} = 1.58\times10^4$ kpc and \nbmc$_{\textrm{,AGN}}=2.4\times10^{-3}$, yields \mbh$_{\textrm{,AGN}}=(1.1\pm0.2)\times10^7~\mathrm{M}_\odot$, where the error reflects the average percentage statistical uncertainty of the \mbh\ $\sim$15 per cent obtained using the GROD05 trend. The same procedure can be used for the other two reference trends. For GXD03, we obtain \mbh$_{\textrm{,AGN}}=(1.6\pm0.4)\times10^7~\mathrm{M}_\odot$ with a percentage uncertainty of 25 per cent, and \mbh$_{\textrm{,AGN}}=(1.2\pm0.5)\times10^7~\mathrm{M}_\odot$ for 4*XTER98 with an uncertainty of $\sim$40 per cent. The \mbh\ uncertainties are obtained from equation~(\ref{eq_Xscal}) plugging $\Gamma\pm 1\sigma$ and \nbmc\ $\pm 1\sigma$ and using the 1$\sigma$ envelope instead of the best fit of the spectral trend described by equation~(\ref{eqjang}). This process is illustrated in Figure \ref{fig:Xrerror} and described in its caption.

\begin{table*}
	\caption{Characteristics of reference sources}		
	\begin{center}
	\begin{tabular}{lcc} 
			\hline
			\hline       
			Name & \mbh & $d$ \\
			& (M$_\odot$) & (kpc)  \\
			\hline
			\gro & $6.3 \pm 0.3$ & $3.2 \pm 0.2$  \\
			\gx & $12.3 \pm 1.4$ & $5.7 \pm 0.8$  \\
			\xte & $10.7 \pm 1.5$ & $3.3 \pm 0.5$  \\
			\hline
		\end{tabular}	
	\end{center}
	\label{table:tabA1}
\end{table*}

\begin{table*}
	\caption{Parametrization of $\Gamma$--\nbmc\ reference patterns}		
	\begin{center}
	\begin{tabular}{lcccc} 
			\hline
			\hline      
			\mcol{Transition} & $A$ & $B$ & $N_\mathrm{tr}$ & $\beta$  \\
			\mcol{(1)} & \mcol{(2)} & \mcol{(3)} & \mcol{(4)} & \mcol{(5)} \\
			\hline
			\gro\ D05 & $1.96\pm0.02$ & $0.42\pm0.02$ & $0.023\pm0.001$ & $1.8\pm0.2$ \\
			\gx\ D03 & $2.13\pm0.03$ & $0.50\pm0.04$ & $0.0130\pm0.0002$ & $1.5\pm0.3$ \\
			\xte\ R98 & $2.96\pm0.02$ & $2.8\pm0.2$ & $0.055\pm0.010$ & $0.4\pm0.1$ \\
			\hline
		\end{tabular}	
	\end{center}
	\begin{flushleft}
		Columns: 1 = reference source spectral transition. 
		2 = parameter that determines the rigid translation of the spectral pattern along the y-axis. 3 = parameter characterizing the lower saturation level of the pattern. 4 = parameter that determines the rigid translation of the spectral pattern along the x-axis. 5 = slope of the spectral pattern. Source: \citet{Gliozzi2011}
	\end{flushleft}
	\label{table:tabA2}
\end{table*}

\begin{table*}
	\caption{Normalization values of $N_\mathrm{BMC,ref}$ as a function of $\Gamma$}		
	\begin{center}
	\begin{tabular}{lcccc} 
			\hline
			\hline      
			\mcol{$\Gamma$} & \nbmc & \nbmc & \nbmc \\
			& \gro\ D05 & \gx\ D03 & \xte\ R98 \\
			\mcol{(1)} & \mcol{(2)} & \mcol{(3)} & \mcol{(4)} \\
			\hline
			1.70 & 0.025 & 0.010 & 0.169 \\
			1.72 & 0.026 & 0.011 & 0.174 \\
			1.74 & 0.027 & 0.012 & 0.179 \\
			1.76 & 0.029 & 0.012 & 0.185 \\
			1.78 & 0.030 & 0.013 & 0.191 \\
			1.80 & 0.032 & 0.014 & 0.197 \\
			1.82 & 0.033 & 0.014 & 0.204 \\
			1.84 & 0.035 & 0.015 & 0.210 \\
			1.86 & 0.037 & 0.016 & 0.217 \\
			1.88 & 0.039 & 0.017 & 0.224 \\
			1.90 & 0.041 & 0.017 & 0.231 \\
			1.92 & 0.045 & 0.018 & 0.239 \\
			1.94 & 0.050 & 0.019 & 0.247 \\
			1.96 & \ldots & 0.020 & 0.255 \\
			1.98 & \ldots & 0.021 & 0.263 \\
			2.00 & \ldots & 0.022 & 0.272 \\
			2.02 & \ldots & 0.023 & 0.281 \\
			2.04 & \ldots & 0.025 & 0.290 \\
			2.06 & \ldots & 0.026 & 0.300 \\
			2.08 & \ldots & 0.029 & 0.310 \\
			2.10 & \ldots & 0.032 & 0.320 \\
			2.12 & \ldots & 0.038 & 0.331 \\
			2.14 & \ldots & \ldots & 0.342 \\
			2.16 & \ldots & \ldots & 0.354 \\
			2.18 & \ldots & \ldots & 0.367 \\
			2.20 & \ldots & \ldots & 0.379 \\
			2.22 & \ldots & \ldots & 0.393 \\
			2.24 & \ldots & \ldots & 0.407 \\
			2.26 & \ldots & \ldots & 0.422 \\
			2.28 & \ldots & \ldots & 0.437 \\
			2.30 & \ldots & \ldots & 0.453 \\
			2.32 & \ldots & \ldots & 0.470 \\
			2.34 & \ldots & \ldots & 0.488 \\
			2.36 & \ldots & \ldots & 0.507 \\
			2.38 & \ldots & \ldots & 0.527 \\
			2.40 & \ldots & \ldots & 0.548 \\
			2.42 & \ldots & \ldots & 0.570\\
			2.44 & \ldots & \ldots & 0.593 \\
			2.46 & \ldots & \ldots & 0.618 \\
			2.48 & \ldots & \ldots & 0.645 \\
			2.50 & \ldots & \ldots & 0.673 \\
			\hline
		\end{tabular}	
	\end{center}
	\begin{flushleft}
		Columns: 1 = photon index $\Gamma$. 2--4 = normalization \nbmc\ values that the three reference sources would have at the same $\Gamma$ as the target AGN. This table provides a simple lookup alternative to using equation~(\ref{eqjang}) once you know $\Gamma$.
	\end{flushleft}
	\label{tab:nbmc}
\end{table*}


\bsp	
\label{lastpage}
\end{document}